\documentclass{aa}
\usepackage{graphicx}
\usepackage{txfonts}
\usepackage{color,xspace}

\newcommand{\equ}[1]{Eq.~\ref{eq:#1}}

\newcommand{\fig}[1]{Fig.~\ref{fig:#1}}

\newcommand{\tab}[1]{Table~\ref{tab:#1}}

\newcommand{\sect}[1]{Sect.~\ref{sec:#1}}

\DeclareMathOperator\erf{erf}
\newcommand{\ssf}[0]{S\'anchez-Salcedo formula\xspace}
\newcommand{\cf}[0]{Chandrasekhar formula\xspace}

\begin{document} 

    \title{Evolution of globular-cluster systems of ultra-diffuse galaxies due to dynamical friction in MOND gravity}

\titlerunning{GCs of UDGs in MOND}

   \author{Michal B\'ilek\inst{1,2,\dag}
   \and Hongsheng Zhao\inst{1,3}
   \and Benoit Famaey\inst{1}
          \and Oliver M\"uller\inst{1}
          \and Pavel Kroupa \inst{4,5}
          \and Rodrigo Ibata\inst{1}
          }
   \institute{$^1$Universit\'e de Strasbourg, CNRS, Observatoire astronomique de Strasbourg, UMR 7550, F-67000 Strasbourg, France\\
	$^2$Nicolaus Copernicus Astronomical Center, Polish Academy of Sciences, Bartycka 18, 00-716 Warsaw, Poland\\
 $^\dag$\email{bilek@asu.cas.cz}\\
    $^3$Scottish Universities Physics Alliance, University of St Andrews, North Haugh, St Andrews, Fife KY16 9SS, UK  \\  
    $^4$Helmholtz-Institut f\"ur Strahlen- und Kernphysik (HISKP), Universit\"at Bonn, Nu{\ss}allee 14-16, D-53115 Bonn, Germany \\
    $^5$Astronomical Institute, Faculty of Mathematics and Physics, Charles University in Prague, V Hole{\v s}ovi{\v c}k{\' a}ch 2, CZ-18000 Praha, Czech Republic}

   \date{Received ...; accepted ...}

% \abstract{}{}{}{}{} 
% 5 {} token are mandatory
 
   \abstract
   {Dynamical friction can be used to distinguish Newtonian gravity and modified Newtonian dynamics (MOND) because it works differently in these frameworks. This concept, however, has yet to be explored very much with MOND. Previous simulations showed weaker dynamical friction during major mergers for MOND  than for Newtonian gravity with dark matter. Analytic arguments suggest the opposite for minor mergers. In this work, we verify the analytic predictions for MOND by high-resolution $N$-body simulations of globular clusters (GCs) moving in isolated ultra-diffuse galaxies (UDGs).}
   {We test the MOND analog of the \cf for the dynamical friction proposed by S\'anchez-Salcedo on a single GC. 
   We also explore whether MOND allows GC systems of isolated UDGs to survive without sinking into nuclear star clusters.}
   { The simulations are run using the adaptive-mesh-refinement code Phantom of Ramses. The mass resolution is $20\,M_\sun$ and the spatial resolution $50\,{\rm pc}$. The GCs are modeled as point masses.}
   { Simulations including a single GC reveal that, as long as the apocenter of the GC is over about 0.5 effective radii, the \ssf works excellently, with an effective Coulomb logarithm increasing with orbital circularity. Once the GC reaches the central kiloparsec, its sinking virtually stops, likely because of the core stalling mechanism. In simulations with multiple GCs, many of them sink toward the center, but the core stalling effect seems to prevent them from forming a nuclear star cluster. The GC system ends up with a lower velocity dispersion than the stars of the galaxy. By scaling  the simulations, we extend these results to most UDG parameters, as long as these UDGs are not external-field dominated. We verify analytically that approximating the GCs by point masses has little effect if the GCs have the usual properties, but for massive GCs such as those observed in the NGC1052-DF2 galaxy, further simulations with resolved GCs are desirable. }
   {}

               
   \maketitle
%
%-------------------------------------------------------------------
\section{Introduction}

Modified Newtonian dynamics, or Milgromian dynamics (MOND; \citealt{milg83b,milg83a}) was introduced four decades ago as a possible explanation for the missing mass or gravity in galaxies not relying on an elusive dark matter component in these stellar systems \citep[see][for a review]{famaey12}. Under MOND theory, rather than adding dark matter, the laws {of} gravity or inertia have to be adjusted in {the} regime of small acceleration $a \ll a_0$, where $a_0$ is on the order of $10^{-10} \, {\rm m}\, {\rm s}^{-2}$ and would constitute a new natural constant. This constant $a_0$ would represent the transition from Newtonian into the so-called deep-MOND regime and could even be related to the cosmological constant \citep[][]{milg99}. The main MOND implication under deep-MOND and spherical symmetry  is that the true acceleration norm $a$ should become $a = \sqrt{a_\mathrm{N} a_0}$, where $a_\mathrm{N}$ is the norm of the Newtonian gravitational acceleration. As an extension of classical gravity, MOND requires a new Poisson equation derived from a Lagrangian. Two such classical MOND Lagrangians have been proposed \citep{bm84, qumond}, while several Lorentz covariant extensions have been proposed over the last two decades; the most recent were {proposed} by \citet{skordis19, skordis20}, which are consistent with the gravitational wave signals.

The MOND theory has celebrated some major achievements on galaxy scales during the last decades. A key prediction of MOND is the tight coupling between the acceleration generated by baryons inside an object and its intrinsic acceleration \citep{milg83b}. About 40 years after its proposition, this coupling has recently been empirically confirmed by high-quality observations of rotation curves of galaxies with stellar masses spanning several orders of magnitude \citep{lelli16,lelli17}. This coupling extends into the regime of low surface brightness galaxies, which were not known at the time of the development of MOND. Strong signs of the external field effect, a unique feature of MOND, have been detected \citep{mcgaugh13a,mcgaugh13b,mcgaugh16b,caldwell17,haghi16, hees16,chae20}.   Simulated galaxy formation under MOND   easily produces exponential galactic disks \citep{wittenburg20}. Furthermore, MOND may provide a solution \citep{zhao13,bil18,banik18b} to 100\,kpc scale phenomena such as the planes-of-satellite problem \citep{pawlowski18,pawlowski20,mueller20planes}. It could also  for instance 
help in shedding light on the origin of large density fluctuations on a 600~Mpc scale \citep{haslbauer20} or on the rapid formation of galaxy clusters \citep{asencio21}.

While MOND is able to account for many aspects of galactic phenomenology -- in particular the detailed shape of the HI rotation curves of spirals,  but also 
X-ray temperature profiles, galaxy-galaxy lensing, and globular cluster (GC) kinematics of most ellipticals \citep{milg12, milg13, samur14, bil19}, this phenomenology can also be understood with appropriate, although fine-tuned, dark matter distributions in the Newtonian context. A profound fundamental difference, however, between MOND and Newtonian dynamics with dark matter halos is the drag experienced by a massive satellite moving through the bath of particles of its host galaxy, which is comprised mostly of stars in the MOND case and stars and dark matter in the Newtonian case. This drag, due to the back-reaction of the wake and angular momentum transfer generated by the massive satellite, is called dynamical friction, and has been not been explored very much in the MOND context. Under MOND, this was pioneered by \citet{ciotti04}, who found with analytical arguments that the dynamical friction timescale is shorter in the deep-MOND regime than with Newtonian gravity including an equivalent dark matter distribution. However, later simulations showed this to be valid only in the regime of small perturbations, while on the contrary, dynamical friction in MOND is rather inefficient when the perturber has a large mass \citep[e.g.,][]{nipoti08, combes14}.  Going further than the mere timescale of dynamical friction, \citet{sanchezsalcedo06} proposed a heuristic formula for the drag force, which we set out in this work to test by $N$-body simulations for the first time. 

{ We focus on the case of GCs moving in and around isolated ultra-diffuse galaxies (UDGs), that is, in galaxies  with very low surface brightnesses and very large radial extents \citep{sandage84,udgs}. Whilst this type of galaxies has been identified for a long time, interest in these galaxies has grown again over the last few years owing to their ubiquity in galaxy groups and galaxy clusters \citep[e.g., ][]{crnojevic14,2015ApJ...807L...2K,munoz15, vandokkum15, udgs, mihos15, yagi16, vanderburg17, venhola17, mueller18,zaritsky19,iodice20}, but also in the field \citep{roman17, leisman17}. In the standard context, many hypotheses have been put forward regarding their possible origin, such as feedback mechanisms that might have made them lose their gas early or expand through dynamical heating \citep[e.g.,][]{Freundlich20} or tidal debris from mergers or tidally disrupted dwarfs \citep[e.g.,][]{Greco2018,Toloba2018, Jiang2019,Carleton2019}. As extreme systems, they are ideal testbeds for various models of galaxy formation. In the context of MOND, irrespective of their formation scenario, such galaxies in the field are very well suited to the study of dynamical friction because they are deep in the MOND regime all the way down to their center. Of course, this is not the case for galaxies residing in an environment in which the external field should dominate their dynamics. The exercise of studying dynamical friction in isolated UDGs is also interesting from the point of view of testing MOND itself. Analytic arguments suggested that GCs of galaxies in the deep MOND regime experience strong dynamical friction and sink quickly { to} the centers of the galaxies \citep{ciotti04,nipoti08}. { Many UDGs} contain many GCs \citep{2018ApJ...862...82L, lim2020}, which might even be exceptionally massive \citep{vandokkum18} and therefore experiencing exceptionally strong dynamical friction. This brings up the question of {whether GCs} could have survived orbiting in (relatively isolated) UDGs since their formation 10\,Gyr ago in MOND, or if MOND { predicts} that such GCs should have sunk { to} the centers of UDGs and formed nuclear star clusters. It is thus important to test the analytic formulas by simulations. The analytic formulas also neglect the gravitational interactions between the individual GCs in the galaxy, which might have slowed down the sinking, as is known from simulations with Newtonian gravity.}

{This paper is organized as follows. In \sect{dynfr} we review the current knowledge of dynamical friction in MOND in more detail and identify the open questions that we later answer by $N$-body self-consistent simulations. The technical details of the simulations and the characteristics of the simulated UDG are described in \sect{setup}. In \sect{stalling} we demonstrate that a single GC does not sink right to the center of the UDG but it experiences core stalling. We further compare the simulation of one GC orbiting the UDG with the prediction of a MOND analog of the Chandrasekhar formula in \sect{ssf}. The evolution of a whole GC system incorporated into our simulated UDG is studied in \sect{manygc}{, where we find the GC system shrinks but the shrinking does not progress to a zero radius}. We then use scaling arguments to {show that the same is expected  for a vast majority of the structural parameters of observed UDGs (as long as they are not external field dominated)} in \sect{scaling}. { In \sect{inter} we investigate analytically how much energy is absorbed {internally in}  GCs during the GC-GC interaction, which is a process that is {not present} in our simulations with GCs modeled as single particles. We find that that our simulations are a good approximation for usual GC masses but less good for the GC masses such as those observed in the NGC1052-DF2 galaxy. The results are summarized  in \sect{conclusions}.}

\section{Dynamical friction}
\label{sec:dynfr}
We first introduce dynamical friction in the context of Newtonian dynamics. Dynamical friction is a process that transfers the orbital energy and angular momentum { of a pair of bodies orbiting each other }into the energy and angular momentum of their constituents.   For example, when a galaxy is orbited by a satellite, the orbital energy of the satellite is transferred to the orbital energy of the stars or dark matter particles of the galaxy. Dynamical friction causes the bodies to merge. It has two dominant sources (see for instance \citealp{mo} or \citealp{just05} for a review). The first is a density wake of the constituent particles behind the object being decelerated. The wake backreacts gravitationally on the orbiting body against the direction of its motion. Supposing Newtonian gravity, the frictional deceleration can in most situations be approximated by the Chandrasekhar formula \citep{chandrasekhar43} as follows:
\begin{equation}
   a_\mathrm{DF, NWT} = \frac{ 2\pi\ln \Lambda G^2\rho m}{\sigma^2 X^2}\left[ \erf(X)-\frac{2X}{\sqrt\pi}\exp\left(-X^2\right)\right].
   \label{eq:chandra}
\end{equation}
In the above equation, $\rho$ stands for the density of the environment of the decelerated body (the enhancement of the density due to the wake is negligible)  and $\sigma$ the local velocity dispersion of the particles of the environment. Next, $m$ denotes the mass of the decelerated body and
\begin{equation}
    X = \frac{v}{\sqrt{2}\sigma},
\end{equation}
where $v$ is the velocity of the decelerated body with respect to its environment. Finally, $\ln\Lambda$ is the Coulomb logarithm. In principle $\Lambda$ is defined as $\Lambda = b_\mathrm{max}/b_\mathrm{min}$, where $b_\mathrm{min}$ is the impact parameter of the background particle whose trajectory is bent owing to the decelerated body by $90\degr$ and $b_\mathrm{max}$  the maximum impact parameter that the background particle can have, that is, in principle the extent of the system. The definition of the Coulomb logarithm is obviously ambiguous and simulations show that its value depends somewhat on the particular problem under consideration \citep[see, e.g.,][for an in-depth criticism of the Coulomb logarithm approach]{Hamilton18}. It is possible to roughly estimate $\ln\Lambda$ from its definition, but if a more precise estimate is needed, a calibration of \equ{chandra} by numerical simulations is required. It turns out that if the Coulomb logarithm is chosen correctly,  \equ{chandra} works very well. The suitable value of $\ln\Lambda$ differs from the default one by a factor of a few. The situations under which the \cf works well were summarized in \citet{mo}. { While the exact value of Coulomb logarithm {has yet to be found} by simulations, { the} \cf is still highly useful without its exact knowledge. { The} \cf gives us a basic understanding of dynamical friction and allows  order-of-magnitude estimates of its {importance}. Formulas based on the \cf are used for example as a basis when we look for fitting formulas for galaxy  merger rates and timescales in simulations \citep{lacey93,jiang08,jiang14,solanes18}.}

The other source of dynamical friction is called  resonant coupling. It happens because the angular momentum of the decelerated body is transferred to the angular momentum of the particles whose orbits are in resonance with the orbit of the  decelerated object. For this reason, a satellite can experience dynamical friction even if it moves outside of the galaxy \citep{lin83,tremaine84}. In such a case no density wake is formed behind the body and \equ{chandra} incorrectly predicts a zero deceleration. In agreement with this second mechanism of dynamical friction, it was found that the effective value of the Coulomb logarithm increases with the circularity of the orbit \citep{chan97}.

In the context of our paper, one more effect is important. It again goes against expectations based on the \cf. It transpires that the satellite can stop its sinking to the center of the host galaxy at a certain radius before reaching the center of the galaxy. This effect is known as core stalling \citep{hernandez98,read06,petts15}. It occurs for galaxies whose gravitational potential is nearly harmonic in the center, that is, having a cored density profile for the case of Newtonian dynamics. In such a potential there are no background particles that would have suitable periods to absorb the energy of the decelerated body. If the density profile is only close to harmonic, the sinking satellite can dynamically heat the central particles and form a core that prevents the sinking again.

Dynamical friction is far less explored in MOND. An important analytic result was obtained by \citet{ciotti04}, assuming the aquadratic formulation of MOND \citep{bm84}. These authros considered a galaxy governed by MOND gravity and the same galaxy in Newtonian gravity with an extra auxiliary rigid gravitational field. The auxiliary field is such that the total gravitational field of the Newtonian galaxy is as in MOND. They found that the ratio of the dynamical friction timescales of the two systems are as follows: 
\begin{equation}
    \frac{t_\mathrm{DF, MOND}}{t_\mathrm{DF, N+AF}}  = \frac{a_0^2}{\sqrt{2}a^2},
    \label{eq:timescalerat}
\end{equation}
where $a$ denotes the typical value of gravitational acceleration in the system. Making use of this and other related equations, they concluded that dynamical friction is more effective in MOND than in Newtonian gravity with live dark matter halos. In spite of this analytic finding, comparative simulations of major mergers of galaxies show the opposite tendency: the merging of galaxies is much faster in Newtonian simulations \citep{nipoti07,tiret07,tiret07c,renaud17}. \citet{combes14}  simulated the orbital evolution of massive gas clumps in young galaxies (clumps containing about 20\% of the baryonic mass of the galaxies) and found that dynamical friction was stronger with Newtonian gravity than in MOND. A resolution of the inconsistency of the analytic calculations and simulations was proposed by  \citet{nipoti08}. These authors verified with a simulation that the analytic estimates work for a small bar rotating in the center of a galaxy. But realistic bars are much larger and take up a significant fraction of the baryonic mass in the central zone. This means that the reservoir of particles to interact with, assumed infinite in the case of the analytic treatment of \citet{ciotti04}, is in reality insufficient to slow down the bar pattern speeds in MOND. In conclusion, { \citet{ciotti04}} suggested that dynamical friction in MOND is not effective if the mass of the perturbers. such as the massive clumps of \citet{combes14}, is comparable to the mass of the whole system because in such a case it is difficult for the remaining ``background'' particles to absorb a large amount of energy.

The MOND results we mentioned so far pertained only to the timescales of dynamical friction. It would be desirable to have a tool to describe the effects dynamical friction in more detail, for example, for predicting the position of the decelerated satellite in time. To this end, \citet{sanchezsalcedo06} proposed an analog to the \cf. They heuristically suggested that  dynamical friction in MOND can be evaluated by multiplying the Newtonian \cf by the ratio of the dynamical friction timescales given by \equ{timescalerat},
\begin{equation}
    a_\mathrm{DF, MOND} = a_\mathrm{DF, NWT}\frac{a_0^2}{\sqrt{2}a^2}.
    \label{eq:ssf}
\end{equation}
This last equation, which we call the \ssf, has never been proven analytically or verified by simulations. We note that even \equ{timescalerat} has only been verified for small galactic bars and not for other orbital configurations of the interacting objects. 

The \ssf has been used for investigating the orbital evolution of the GCs of the Fornax dwarf galaxy \citep{sanchezsalcedo06, angus09}.  Analytic estimates predicted a fast sinking of GCs to the centers of dwarf galaxies, where the GCs should  merge and form a nuclear star cluster \citep{ciotti04,sanchezsalcedo06,nipoti08}. The presence of GCs and the absence of a nuclear star cluster in Fornax were initially claimed to contradict MOND, but it was later found that the problem can be solved if we assume that the GCs were formed further away from the galaxy than where they are observed today \citep{angus09}. Moreover, the applicability of the analytic formula  has never been verified by simulations for the case of GCs orbiting a galaxy.

The discovery of a large number of UDGs (e.g., \citealt{sandage84,impey88,crnojevic14,2015ApJ...807L...2K,vanderburg17,mueller18,habas20,iodice20}) brings up the question of survival of GC systems again. These diffuse objects, with a surface brightness fainter than $\sim 24$ in the $g$ band and an effective {radius} larger than 1.5\,kpc, can contain a large number of GCs \citep{beasley16,vandokkum16,vandokkum18,2020A&A...640A.106M}. The fact that UDGs are deep-MOND objects \citep{2019A&A...623A..36M} and the GCs of some of these objects seem to be exceptionally massive \citep{vandokkum18} indicates that the GC systems of UDGs can be affected strongly by dynamical friction.

Newtonian simulations have already been performed to investigate the evolution of the GC system of the UDG NGC\,1052-DF2 (DF2 hereafter) by \citet{duttachowdhury19}. This galaxy has been claimed to have a system of unusually massive GCs \citep{vandokkum18} if it lies at the distance of 20\,Mpc \citep{danieli20}, which we assume hereafter (see e.g., \citealt{trujillo19} or \citealt{haghi19} for a discussion of the effect of the distance on the properties of DF2). The galaxy hosts ten confirmed GCs, all of which have stellar masses between $10^5$ and $10^6\,M_\sun$. They contain a non-negligible fraction of the total stellar mass of the galaxy. The simulations by \citet{duttachowdhury19} captured several effects that do not readily follow from the Chandrasekhar formula or its modified-gravity analogs -- the core stalling and the interactions of GCs among each other (the so-called dynamical buoyancy). Such effects have never been  included in the studies of the orbital decay of GCs in MOND.

There are many open questions regarding dynamical friction in MOND. {In this paper, we wanted to investigate several of them using self-consistent $N$-body simulations. First, we wanted to explore the precision of the \ssf by studying the motion of a single GC in a UDG. We wanted to learn whether MOND really predicts that the GC will sink in the center of the galaxy over a few {gigayears}, as following the \ssf. Then we were interested whether the answer will change if we include in the simulation several GCs that interact with each other. The arising dynamical buoyancy could possibly slow down the sinking. These simulation would eventually address whether MOND predicts that GCs of UDGs should have merged and formed nuclear stars clusters in a few gigayears after the formation of the GCs, if the GCs were initially distributed around the galaxy at similar distances as observed today.}

Our simulations were inspired by the DF2 galaxy and its possibly overmassive GC system. It is a typical UDG in terms of stellar mass, size and S\'ersic index \citep{2020MNRAS.491.1901H}. 
We stress that our simulations are not intended to describe DF2 itself because in MOND this galaxy is subjected to a strong external field effect \citep{famaey18,kroupa18,haghi19}. Thus, MOND predicts DF2 to behave in the Newtonian way. Our present investigations pertain to the study of DF2 analogs that would be far from a massive neighbor and hence fully unaffected by the external field effect.

\begin{table}
\caption{Parameters of the simulations.}
\label{tab:sim}
\begin{tabular}{l|l}
\hline\hline
Property/parameter  & Value                \\
\hline
Number of particles & $10^{7}$\\
Mass of star particle & $20\,M_\sun$ \\
\texttt{levelmin} & 7\\
\texttt{levelmax} & 11 \\
\texttt{boxlen} & 100\,kpc \\
Maximum resolution & 0.049\,kpc \\
\texttt{mass\_sph}  & 1 \\
\texttt{m\_refine} & $200\,M_\sun$ for every level \\
\hline
\end{tabular}
\end{table}

\begin{figure}
        \resizebox{\hsize}{!}{\includegraphics{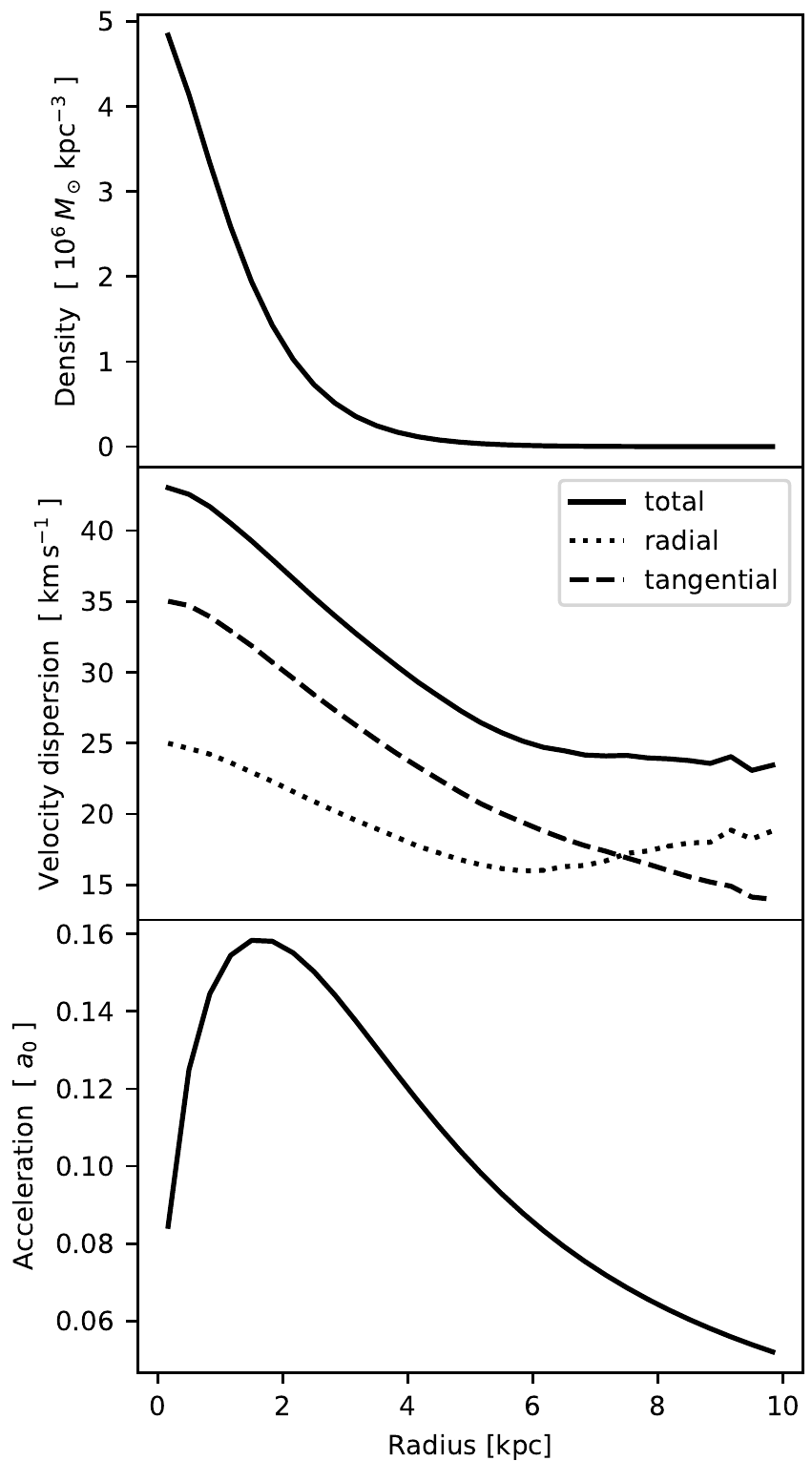}}
        \caption{Properties of the simulated galaxy just before entering GCs. Top: Radial profile of density. Middle: Radial profile of velocity dispersion, separately for the full velocity dispersion (full line), velocity dispersion in the radial direction (dotted line), and in the tangential direction (dashed line). Bottom: Radial profile of gravitational acceleration in the units of the MOND { constant $a_0$}. }
        \label{fig:ics}
\end{figure}

%%%%%%%%%%%%%%%%%%%%%%%%%%%%%%%%%%%%%%%%%%%%%%%
%%%%%%%%%%%%%%%%%%%%%%%%%%%%%%%%%%%%%%%%%%%%%%%
\section{Setup of the simulations}
\label{sec:setup}
In this work, we assumed the QUMOND formulation of MOND \citep{qumond}. If the distribution of matter $\rho$ is given, then the gravitational field can be obtained by solving the QUMOND equivalent of the Poisson equation:
\begin{equation}
    \Delta\phi=\vec \nabla\cdot[\nu(|\vec\nabla\phi_N|/a_0)\vec\nabla\phi_N],
\end{equation}
where 
\begin{equation}
    \Delta\phi_N= 4\pi G \rho    
\end{equation}
is the standard Poisson equation for the Newtonian potential $\phi_N$. We employed the so-called simple form of the interpolation function $\nu$,
\begin{equation}
    \nu(x) = \frac{1 + (1 + 4x^{-1})^{1/2}}{2},
    \label{eq:nu}
\end{equation}
and the value of the MOND acceleration constant of $a_0 = 1.12\times 10^{-10}$\,m\,s$^{-2}$, which is consistent with the measurements of \citet{mcgaugh16}. 

The simulations were performed using the Phantom of RAMSES code \citep{por}, which is a QUMOND adaptation of the popular adaptive-mesh-refinement code RAMSES \citep{ramses}. For the simulation described below, unless specified otherwise we used the computational parameters summarized in \tab{sim}.
We simulated the galaxy with $10^7$ particles of $20\,M_\sun$. { Each GC was represented by a point mass. For reference, the overmassive GCs of DF2 were estimated to have effective radii of around 8\,pc \citep{vandokkum18}.} The grid was refined if the mass in the cell exceeded $200\,M_\sun$, that is, if it contained at least ten stellar particles or one GC. The maximal spatial resolution was 49\,pc.

\begin{figure*}
        \centering
        \includegraphics[width=17cm]{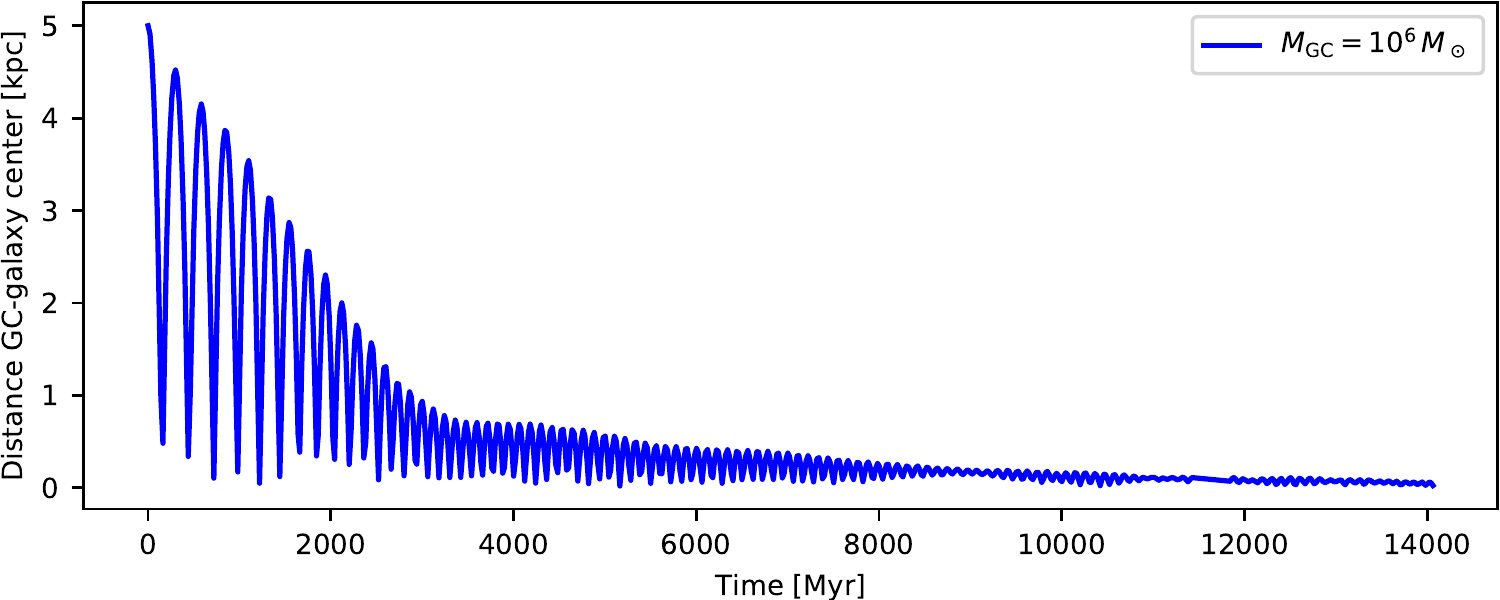}
        \caption{Orbital decay of the GC in the fiducial model. The decay slows down substantially once the apocentric distance of the GC reaches about 0.5-1\,kpc. } 
        \label{fig:fiducial}
\end{figure*}

\begin{figure*}
        \centering
        \includegraphics[width=17cm]{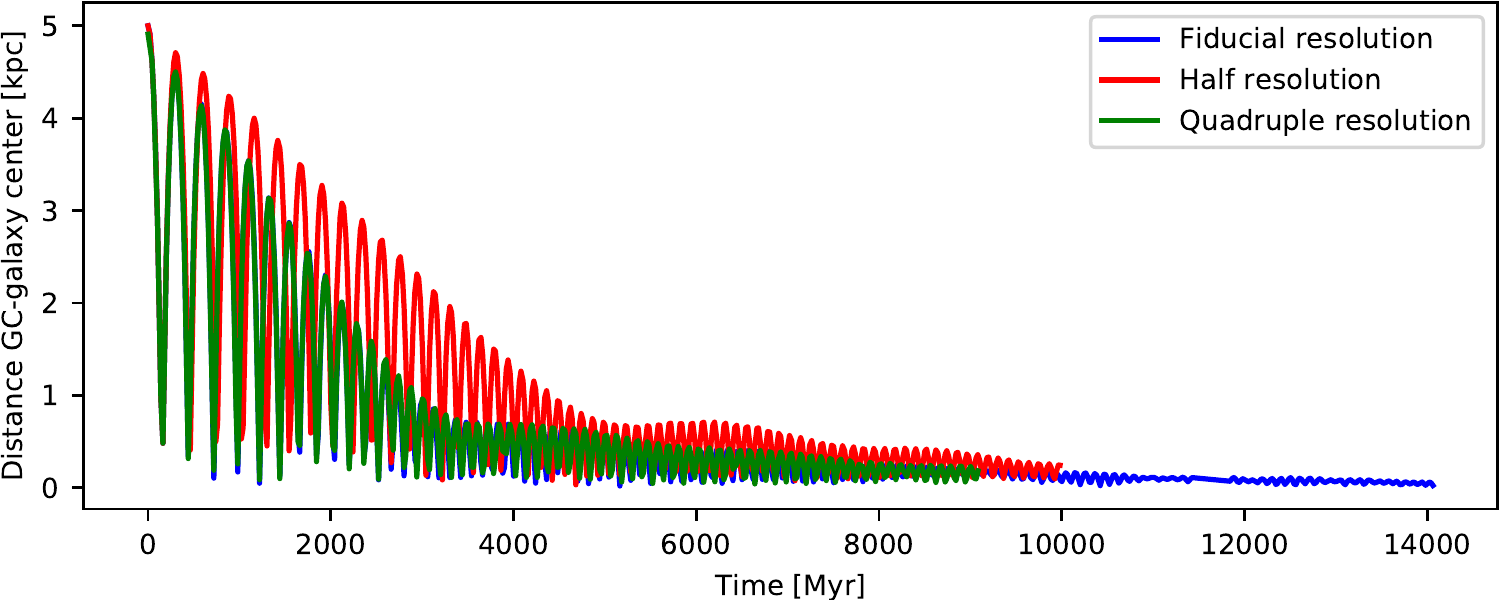}
        \caption{Demonstration of the effect of changing of the spatial resolution in the fiducial model on the orbital decay of the GC. The core stalling phase is present for all three choices of resolution.} 
        \label{fig:resolution}
\end{figure*}

\begin{figure*}[h!]
        \centering
        \includegraphics[width=17cm]{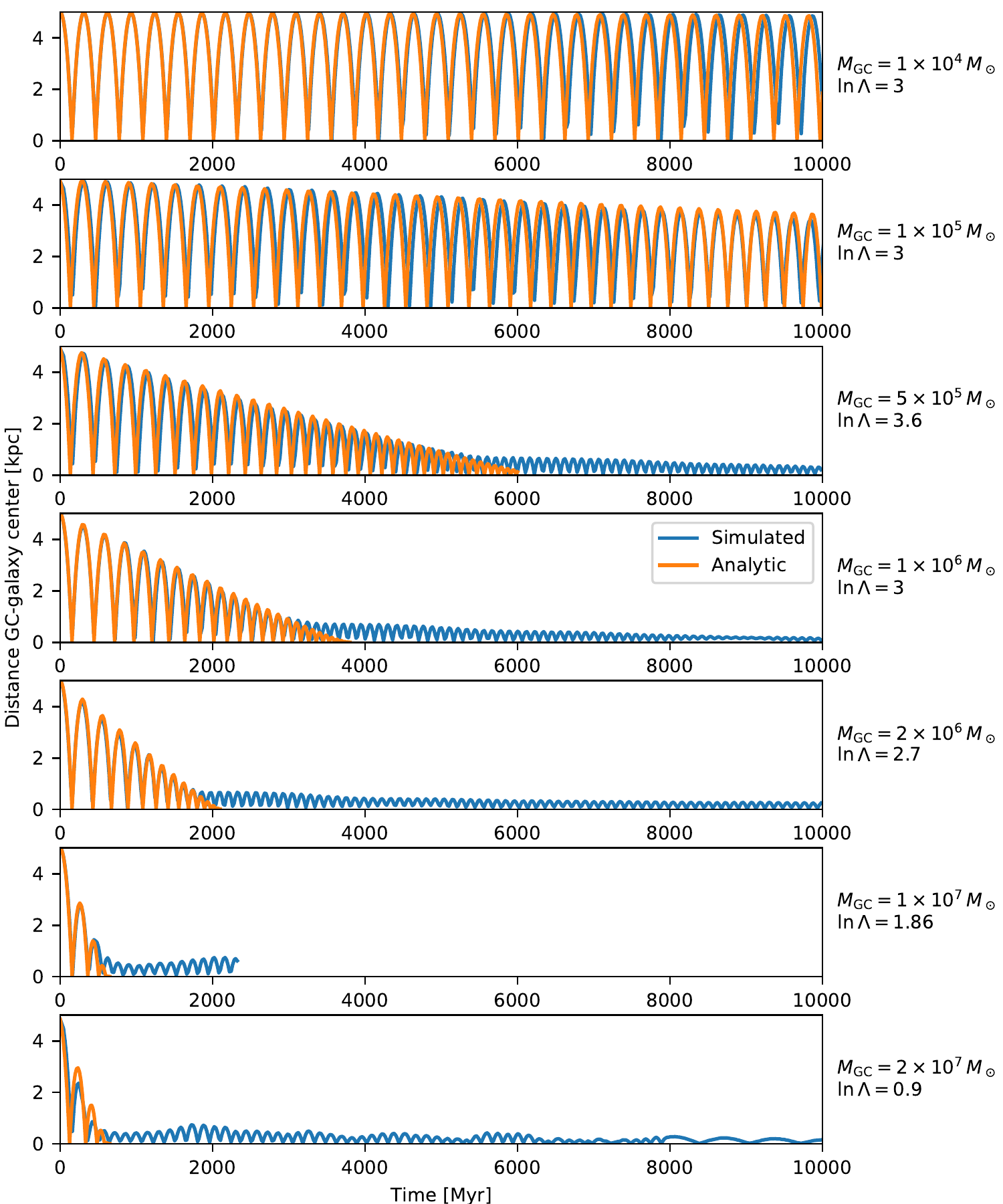}
        \caption{Testing the precision of the \ssf for dynamical friction in MOND (\equ{ssf}). Blue lines: Evolution of the GC-galaxy distances in $N$-body simulations. Orange lines: The same but calculated analytically employing the \ssf. Each panel corresponds to another mass of the GC under consideration. The GCs in all simulations or calculations started with the same initial conditions, that is, starting from rest with respect to the galaxy and at a distance of 5\,kpc from the galaxy center. To the right of each figure, the corresponding mass of the GC and the value of the Coulomb logarithm that is needed to be entered in the \ssf to mimic the result of the simulation are stated.  } 
        \label{fig:massescomp}
\end{figure*}

\begin{figure*}
        \centering
        \includegraphics[width=17cm]{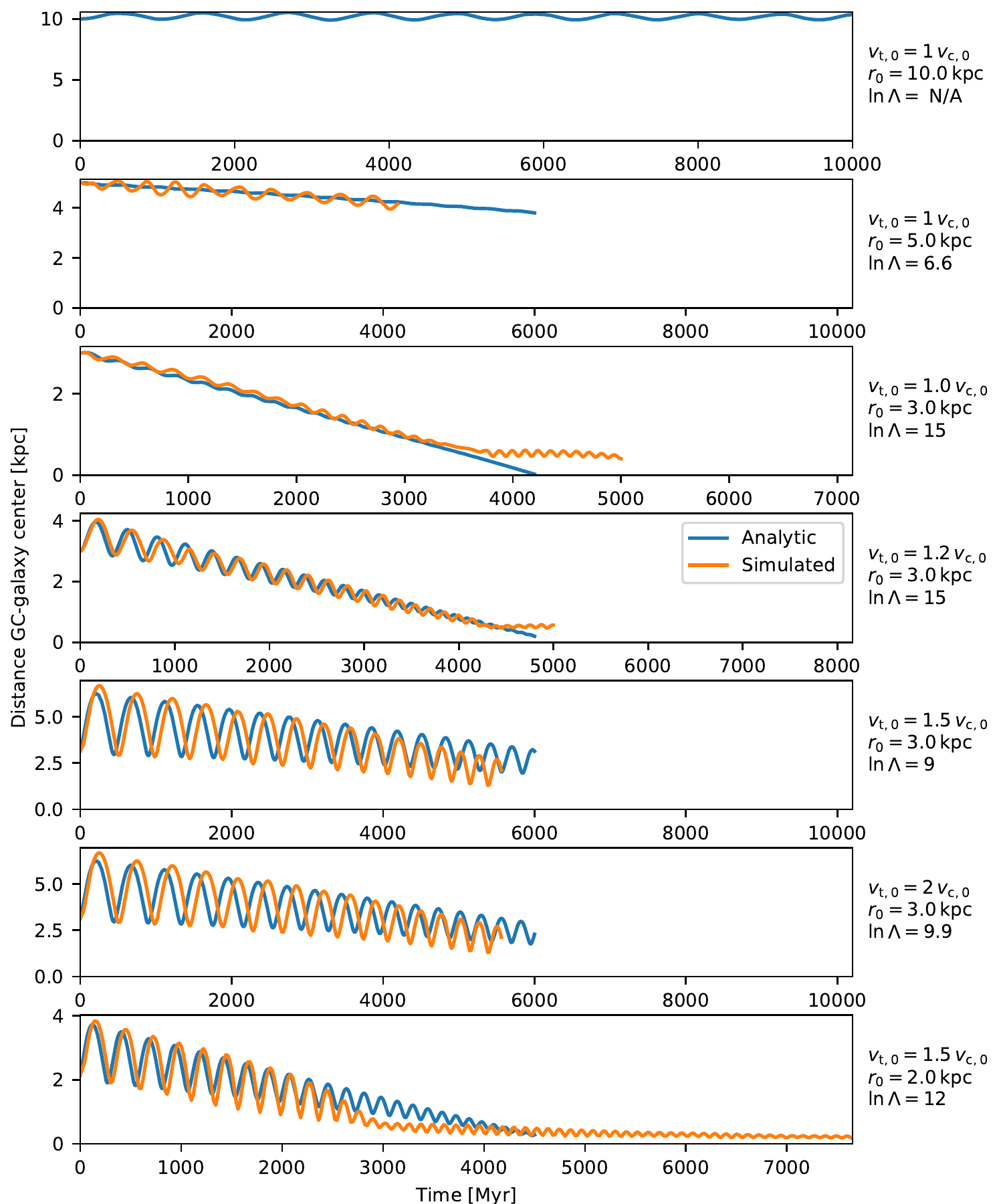}
        \caption{As in \fig{massescomp}, but varying orbital parameters of the GC. The GC always had the mass of $10^6\,M_\sun$. The {first lines of the notes on the right of the panels indicate} the initial tangential velocity of the GC with respect to the galaxy (the radial velocity was zero). It is given in the units of the local circular velocity. The second row indicates the initial galactocentric distance and the third the value of the  Coulomb logarithm in the \ssf that provides the best match of the simulation and analytic calculation.  } 
        \label{fig:ecccomp}
\end{figure*}

\begin{figure*}
        \centering
        \includegraphics[width=17cm]{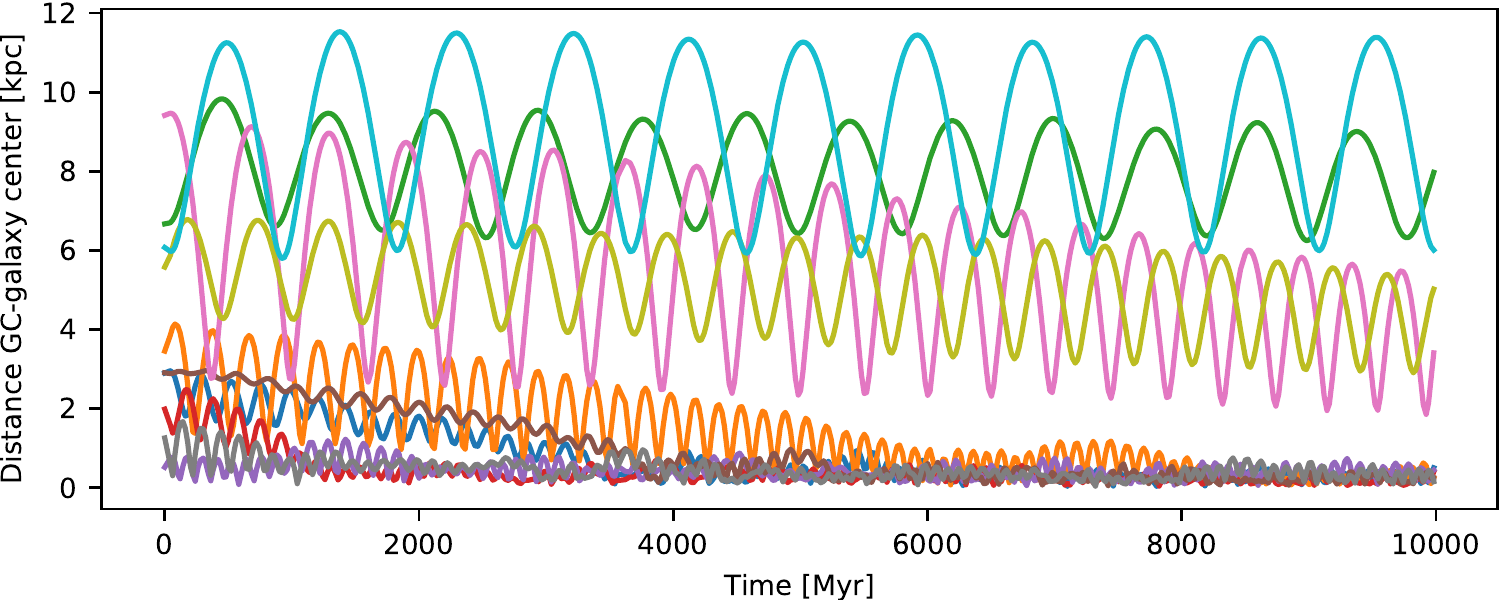}
        \caption{Example of evolution of galactocentric radii of GCs in a simulation with 10 GCs. The masses of the GCs are set as observed for the GCs of the DF2 galaxy (first column of \tab{gcmass}).} 
        \label{fig:multigc}
\end{figure*}

We initialized the density of the simulated galaxy as a S\'ersic sphere that has the parameters observed for DF2, that is, a stellar mass of $2\times10^8\,M_\sun$, effective radius of 2\,kpc, and S\'ersic index 0.6 \citep{2018Natur.555..629V}. The galaxy was modeled with $10^7$ particles, each of a mass of $20\,M_\sun$. Each particle was given a velocity drawn from a three-dimensional isotropic Gaussian distribution. The local velocity dispersion was determined by solving the spherically symmetric Jeans equation. Before introducing any GCs into the simulation, the galaxy was let to virialize for 10\,Gyr (over 70 free-fall times from the distance of two effective radii to the center of the galaxy). After this period, a S\'ersic fit of the density yields an effective radius of 1.9\,kpc and a S\'ersic index of 0.7.
The radial profiles of density, velocity dispersion, and gravitational acceleration after the virialization are plotted in \fig{ics}. The inner $\sim$7\,kpc of the galaxy became dominated by stars on tangential orbits, while the more distant regions are inhabited by stars preferentially on radial orbits.

%%%%%%%%%%%%%%%%%%%%%%%%%%%%%%%%%%%%%%%%%%%%%%%
%%%%%%%%%%%%%%%%%%%%%%%%%%%%%%%%%%%%%%%%%%%%%%%

\section{Core stalling:\ Demonstration on the fiducial model}
\label{sec:stalling}
We begin presenting our results with the simulation that we hereafter refer to as the fiducial. It starts with one GC placed at the distance of 5\,kpc from the center of the galaxy such that the two objects are at rest with respect to each other. The GC is modeled as a single particle with a mass of $10^6\,M_\sun$, mimicking some of the most massive GCs of DF2. Figure~\ref{fig:fiducial} shows the time evolution of the distance of the GC from the center of the galaxy. We can note that the apocentric distances initially decrease at an approximately constant rate, but at a time of about 3\,Gyr the orbital decay slows down dramatically. This effect has also been noted in the Newtonian simulations of DF2 of \citet{duttachowdhury19}, who attribute the effect to the core-stalling mechanism. 

Nevertheless, our simulation is not Newtonian. It is important to verify that the observed core stalling is not due to numerical errors. We can immediately check that the apocentric radius of the GC at the beginning of the stalling phase, at about 0.5\,kpc, is an order of magnitude higher than  the maximum spatial  resolution of the simulation, 0.049\,kpc. In addition, we run extra simulations with a higher and lower spatial resolution to demonstrate the convergence of the simulation for our default parameters. The result is shown in \fig{resolution}. The orbits of the GCs in the fiducial model and that with quadruple resolution nearly coincide. The agreement is not that tight when comparing the default and half resolution simulations, but they still agree well on the apocentric distance of the GC in the stalling phase. This demonstrates that the default resolution is a good balance between precision and computing demands. In total, we found no hints that the observed core stalling would be caused by an insufficient resolution of the simulation.

%%%%%%%%%%%%%%%%%%%%%%%%%%%%%%%%%%%%%%%%%%%%%%%
%%%%%%%%%%%%%%%%%%%%%%%%%%%%%%%%%%%%%%%%%%%%%%%
\section{Testing the \ssf for dynamical friction in MOND}
\label{sec:ssf}
The  \ssf (\equ{ssf}) is supposed to be applicable to objects in the deep-MOND regime (accelerations lower than $a_0$), which is the case of our UDG (see \fig{ics}). We investigated whether the orbital decay of a GC can be modeled analytically by solving the {following} equation of motion:
\begin{equation}
    \vec{a} = \vec{a_\mathrm{grav}}(|\vec{r}|)+\vec{a_\mathrm{DF}}(|\vec{r}|,|\vec{v}|; \Lambda),
    \label{eq:em}
\end{equation}
for a suitable value of the Coulomb logarithm $\Lambda$, where $a_\mathrm{grav}$ is the gravitational acceleration given by \equ{mond} and $a_\mathrm{DF}$ the acceleration caused by dynamical friction according to the \ssf. We assumed that the mass of the GC is negligible compared to the mass of the galaxy, such that we can treat the GC as a test particle. A similar approach was used in \citet{sanchezsalcedo06} or \citet{angus09}. Solving such an equation is of course much faster than running a simulation.

When solving \equ{em}, we made use of the fact that the radial acceleration of a test particle in the gravitational field of a spherically symmetric object in QUMOND can be evaluated as
\begin{equation}
    a_\mathrm{grav} =  a_\mathrm{grav, N}~\times \nu\left(\frac{\left| a_\mathrm{grav, N}\right|}{a_0}\right), 
    \label{eq:mond}
\end{equation}
where $a_\mathrm{grav,N}$ is the radial acceleration expected from Newtonian dynamics \citep{qumond}. In order to evaluate $a_\mathrm{grav, N}$, we modeled the mass profile of the galaxy analytically as a S\'ersic sphere with the post virialization parameters of our stimulated galaxy described in \sect{setup}. The density of the S\'ersic sphere was calculated by exploiting the analytic approximations by \citet{sersdeproj}, updated in \citet{sersdeprojupdate}. When solving numerically the equation of motion, we also added 1\,pc to the distance of the GC from the center of the UDG to avoid numerical problems close to the center of the galaxy.

\subsection{Varying the mass of the GC}
First, we explored the role of the mass of the GC, keeping the initial orbital configuration fixed. At the beginning of each simulation, we again placed the GC at the distance of 5\,kpc from the center of the UDG such that the two objects had a zero relative velocity. We only varied the mass of the GC in each simulation. The resulting evolution of distance of the GCs from the galaxy center in each simulation are shown in \fig{massescomp} with the blue lines. The mass of the GC in each simulation is indicated next to the corresponding panel of the figure.

The simulations were compared to the analytic solutions of the model (\equ{em}). They are indicated in  \fig{massescomp} by the orange lines. The values of the Coulomb logarithm were fitted manually to obtain a balance between matching  the instants of apocenters and matching of the overall slopes of the curves. The fitted values are again shown next to the corresponding panels of the figure. All realistic values for the mass of the GC require a value of $\ln\Lambda \simeq 3$ for these radial orbits. This value has already been proposed by \citet{sanchezsalcedo06}.

\begin{table}
\caption{Masses of GCs in the many GC simulations in units of $10^5\,M_\sun$.}
\label{tab:gcmass}
\begin{tabular}{lll}
\hline\hline
& DF2-like  & Standard               \\
\hline
& 15.0  &  8.5  \\
& 9.6  &  4.8  \\
& 8.0  &  3.5  \\
& 7.3  &  2.7  \\
& 6.6  &  2.1  \\
& 6.6  &  1.7  \\
& 5.5  &  1.3  \\
& 5.0  &  1.0  \\
& 4.2  &  0.7  \\
& 3.8  &  0.4  \\ \hline
Mean: & 7.2 & 2.7 \\
\hline
\end{tabular}
\end{table}

\subsection{Varying the orbital eccentricity of the GC}
We also explored how the \ssf works for non-radial orbits. The mass of the GC was fixed at $10^6\,M_\sun$. The GC was always initiated as moving in the tangential direction with respect to the center of the galaxy. We specified the orbital initial conditions by the initial distance of the two objects and their relative tangential velocity in units of the circular velocity at the given radius evaluated from \equ{mond}.

The results are presented in \fig{ecccomp}, where each panel represents one simulation. The text lines on the right of each panel give the initial tangential velocity, initial distance of the GC from the galaxy, and the fitted value of the Coulomb logarithm, respectively. The main conclusion of these experiments is the  finding that the effective value of the Coulomb logarithm is higher than in the case of the radial orbits as it ranges between the values of 6.6 and 15.  In the case of the GC moving on a circular orbit with a radius of 10\,kpc = 5.2\,$R_\mathrm{e}$, there is no clear hint of orbital decay in the simulation because of the low density of background particles. The analytic model gives good agreement with this simulation for any reasonable value of the Coulomb logarithm; we tested the range between 0 and 50.

\begin{figure*}[t!]
        \centering
        \includegraphics[width=17cm]{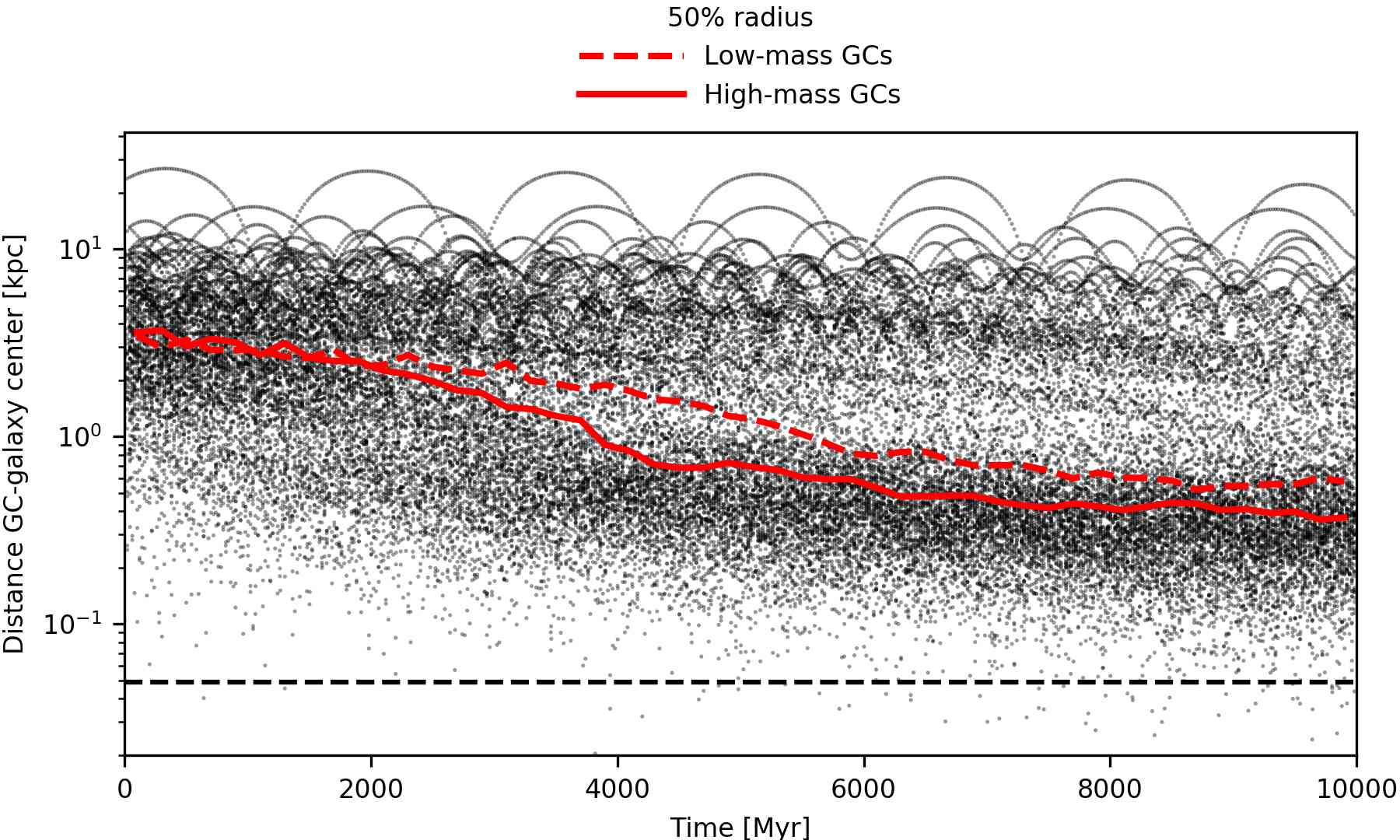}
        \caption{Evolution of the distances of GCs from the center of the UDG. The black points show combined data from 10 random realizations of a UDG orbited by 10 GCs. The masses  of the GCs in each simulation mimic the measured masses of the GCs of the DF2 galaxy (see the first column of \tab{gcmass}). The horizontal black dashed line indicates the maximum resolution of the simulation. The red lines indicate the median galactocentric distance of the GCs at the given time. This was derived separately for the high-mass half of the GCs (full red line) and for the low-mass half of the GCs (dashed red line). } 
        \label{fig:rtall}
\end{figure*}

\begin{figure*}[h!]
        \centering
        \includegraphics[width=17cm]{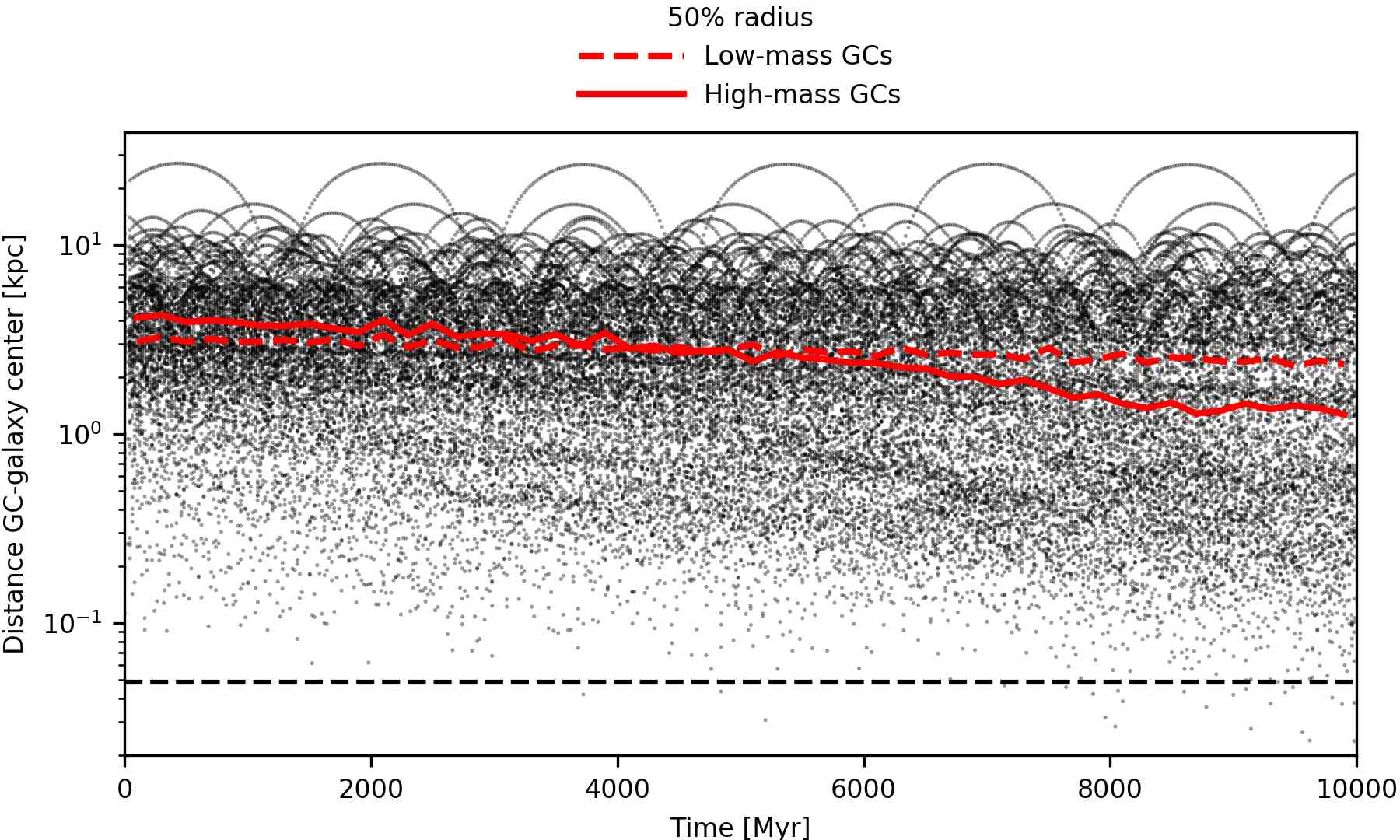}
        \caption{The same as in \fig{rtall} but with GC masses set according to the standard GCLF. } 
        \label{fig:rtallgclf}
\end{figure*}

\begin{figure*}[h!]
        \centering
        \includegraphics[width=17cm]{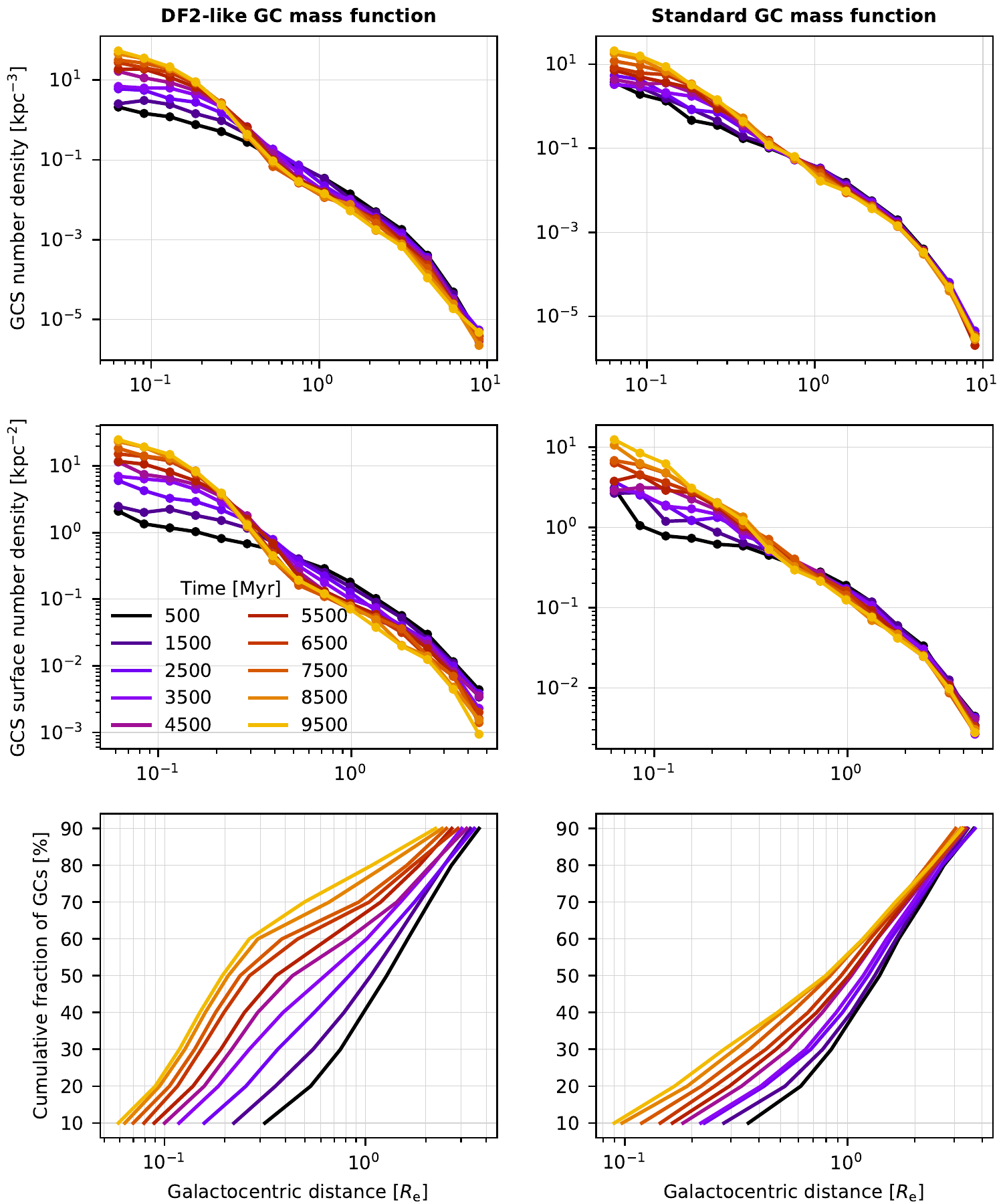}
        \caption{Evolution of GC systems. Left column: The simulations with 10 GCs that have the masses of those observed in DF2 (left column of \tab{gcmass}). Simulations of 10 random realizations of the system were combined in these plots. Galactocentric distance is given in units of the effective radius of the galaxy ($R_\mathrm{e}= 1.9$\,kpc). Right column: The same but for GC masses set according to the standard GCLF (right column of \tab{gcmass}). Top row: Profiles of number density of GCs as a function of  galactocentric distance. The individual lines correspond to different times since the start of the simulation. Middle row: Evolution of the profile of the surface number density of GCs. Bottom row: Evolution of the radial profile of the cumulative fraction of GCs.  } 
        \label{fig:gcsprofs}
\end{figure*}

%%%%%%%%%%%%%%%%%%%%%%%%%%%%%%%%%%%%%%%%%%%%%%%
%%%%%%%%%%%%%%%%%%%%%%%%%%%%%%%%%%%%%%%%%%%%%%%
\section{Evolution of the density profile of the GC system of a UDG}
\label{sec:manygc}
The properties of the GC systems of UDGs are still a matter of debate. We considered here two types of GC systems. In one, the GCs had the high masses of the GCs of DF2. In the other the masses of GCs were consistent with the standard globular cluster luminosity function (GCLF) of dwarf galaxies. The standard GCLF mass function seems to be more common for observed UDGs at least in the Fornax and Virgo clusters \citep{2019MNRAS.484.4865P,lim2020}. We generally obtained different results for these two types of GC systems.

\subsection{DF2-like GC mass function}
\label{sec:df2mf}
We added into the galaxy 10 GCs that have masses as inferred for DF2 \citep{duttachowdhury19}, see \tab{gcmass}. Their spatial distribution was inspired by the distribution of the GCs of DF2.  \citet{duttachowdhury19} found that the GC system of this galaxy can be described by a S\'ersic sphere  of the index 1 whose effective radius is 1.3 times the effective radius of the galaxy.

The initial conditions for the GC system were prepared in the following way. We initially drew the positions of the GCs randomly from a distribution matching the observed S\'ersic profile. The GCs were first assigned velocities in the same way as we did for the initial velocities of the particles of the galaxy (see \sect{setup}). It was then necessary to let the initial conditions for the GCs virialize while avoiding  GC-GC interactions and the dynamical friction on the particles of the galaxy. To achieve this, we extracted the virialized mass profile of the galaxy (see \sect{setup}), calculated its spherically symmetric gravitational potential, and integrated the equation of motion of each GC individually in this rigid potential for 10\,Gyr. The resulting positions and velocities of the GCs were then used as the initial conditions for the GCs in the self-consistent simulation. Contrary to \citet{duttachowdhury19}, we did not strive to assign the GCs the observed radial velocities or the observed projected galactocentric distances because we aimed to investigate only a general UDG similar to DF2.

We ran ten simulations, each for a different random realization of the initial conditions for the GCs. The time evolution of galactocentric distances of the GCs in one of those simulations is plotted in \fig{multigc}. This plot shows that the motion of individual GCs are influenced by the others, particularly those near the center of the galaxy.  The positions of the GCs were recorded in the simulations approximately every 20\,Myr. The recorded galactocentric distances of the GCs in all simulations are plotted together in \fig{rtall}. This figure shows how the radial profile of the GC density evolves on average. The GCs tend to gather in the central 1\,kpc of the galaxy after a few {gigayears}. The profile of GC density gets initially depleted at around a radius of 3\,kpc (1.5\,$R_\mathrm{e}$ of the galaxy), but later the depleted region starts extending toward lower and higher radii. In order to get more quantitative results, we inspected the half-number radius of the GCs. We divided the sample of GCs in halves according to the masses of GCs. The full line in the figure indicates the half-number radius of the high-mass group of GCs and the dashed line is the same for the low-mass group. The figure shows that the high-mass GCs settled faster because of mass segregation. The GCs did not sink right to the very center, likely again because of the core stalling mechanism. The horizontal dashed line in the figure indicates the highest spatial resolution of the simulation. It is again substantially smaller than the radius where the stalling occurs. { It} can be noted that the core stalling phase establishes after about half of the typical age of a GC. The fact that the spatial distribution of GCs of the DF2 galaxy does not have a core suggests that it has spent the majority of its life under the influence of a strong EFE imposed by its massive neighbor NGC\,1052.

We divided this dataset into 1\,Gyr time intervals. The data points in each interval were converted to radial profiles of GC number density, surface number density, and cumulative fraction of GCs inside a given radius. The results are shown in the left column of \fig{gcsprofs}. Each line in these plots corresponds to a given time interval. The centers of the time intervals are indicated in the legend according to the color key. These plots can be compared to observations.

In order to compare a galaxy to our simulation with a single parameter, we propose hereafter an observational concentration parameter (not yet used elsewhere to the best of our knowledge) defined as
\begin{equation}
R = \frac{N(<1R_\mathrm{e})-N(<0.5R_\mathrm{e})}{N(<1R_\mathrm{e})}, 
\label{eq:r}
\end{equation}
where the symbol $N(<x)$ denotes the number of GCs projected inside a circle of radius $x$. The time evolution of the $R$ parameter in our simulations is shown in \fig{revol}. For example, for the real DF2 galaxy, we have $N(<1R_\mathrm{e})=6$ and $N(<0.5R_\mathrm{e})=2$ \citep{duttachowdhury19} and therefore $R = 0.67$, such that this galaxy does not show any hint of a central concentration. 

{ The distribution of velocities of GCs also evolves. Figure~\ref{fig:veldisp} shows that the line-of-sight velocity { dispersion} of the whole GC system decreases by about one-third during the simulated period of 10\,Gyr. The line-of-sight velocity dispersion measured from the velocities of the five GCs projected the closest to the center, decreases even more, dropping from about 23\,km\,s$^{-1}$ at the beginning of the simulation to 8\,km\,s$^{-1}$ at its end. It is interesting that the velocity dispersion of this inner part of the GC system stabilizes after the first 6\,Gyr of the simulation, likely due to the core stalling. The radial profiles of the line-of-sight velocity dispersion of the GC system at the beginning and end of the simulation are compared in the top panel of \fig{siglosprof}. This shows that the drop of velocity dispersion happened mostly in the inner part of the GC system. The same plot also represents the line-of-sight velocity dispersion profiles of the stars of the galaxy at the beginning and end of the simulation. In contrast with the GCs, the central velocity dispersion of the stars increased with time. There was also a mild increase in velocity dispersion of the stars in the outer parts of the galaxy. We attributed these changes to energy equipartition of stars and GCs. The radial profiles of the line-of-sight velocity dispersions of stars and GCs are not the same even at the beginning of the simulation. This is because of a different density profile of the stars and GCs. A different profile of the anisotropy parameter can also play a role.}

\begin{figure}
        \resizebox{\hsize}{!}{\includegraphics{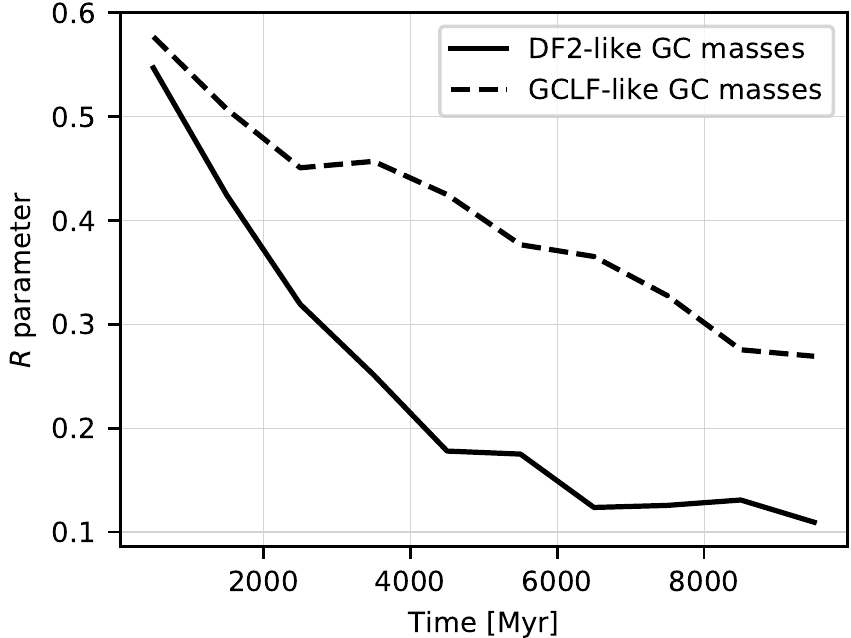}}
        \caption{ Time evolution of the concentration parameter $R$ of the GC system defined by \equ{r}. The two lines correspond to the two considered distributions of the masses of GCs (see \tab{gcmass}).} 
        \label{fig:revol}
\end{figure}

\begin{figure}
        \resizebox{\hsize}{!}{\includegraphics{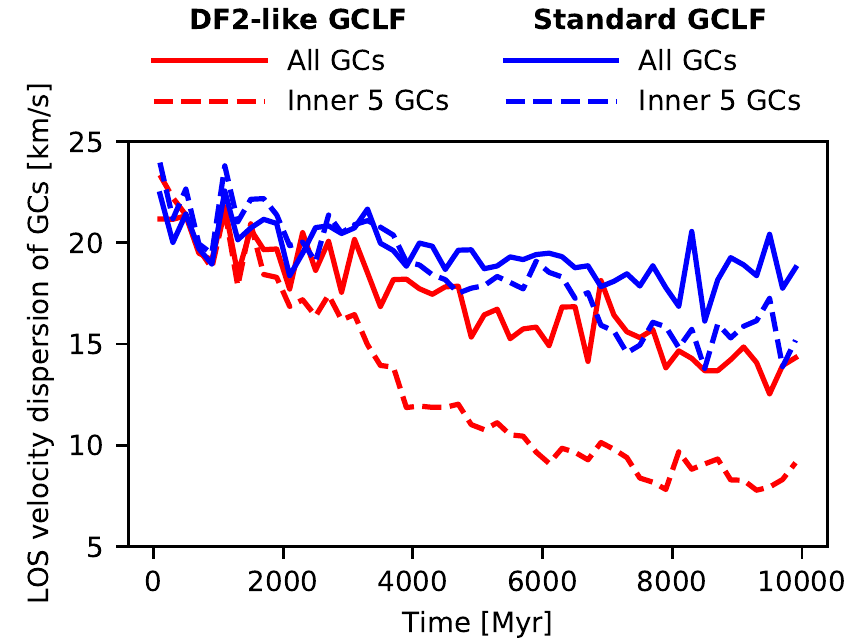}}
        \caption{Line-of-sight velocity dispersions of the GC systems in our simulations as functions of time.}
        \label{fig:veldisp}
\end{figure}

\begin{figure}
        \resizebox{\hsize}{!}{\includegraphics{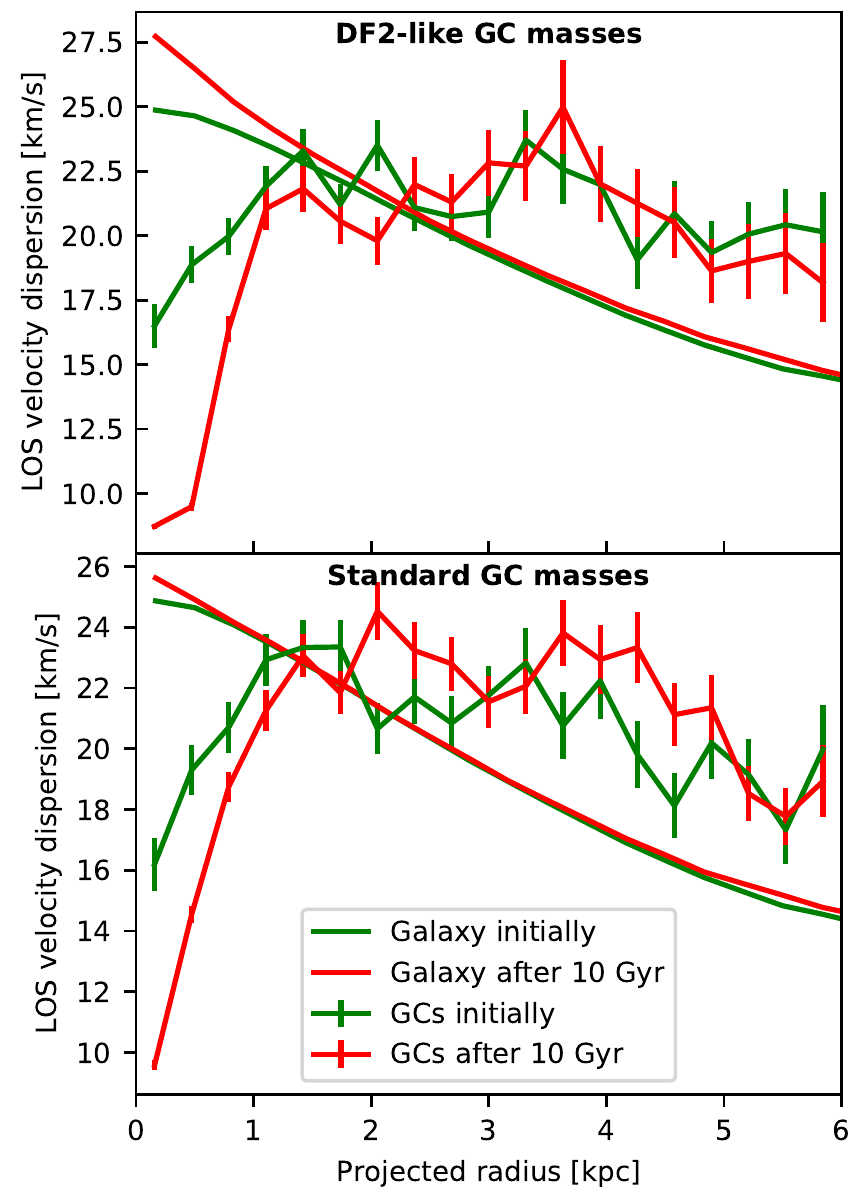}}
        \caption{Radial profile of the velocity dispersion profile of the stars  and of the GC system of the simulated galaxy. The curves correspond to the beginning and end of the simulations, as indicated in the legend. The vertical lines indicate the $1\sigma$  uncertainty limits.}
        \label{fig:siglosprof}
\end{figure}

\subsection{Standard GC mass function}
We repeated the same procedure as above but substituting the masses of the GCs with values according to a GCLF typical for dwarf galaxies \citep{1996AJ....112..972D,2009MNRAS.392..879G}. Other than the GC masses, the initial conditions of the GCs remained the same. The distribution of absolute magnitudes of GCs in a galaxy can generally be described by a Gaussian distribution. It turns out that the mean of the  distribution is common for nearly all galaxies. Therefore the peak of the GC luminosity distribution is being used as a distance indicator \citep{2012Ap&SS.341..195R}. The width of the Gaussian distribution depends on the mass of the galaxy \citep{jordan07,villegas10}. For UDG galaxies in the Virgo Cluster, \citet{lim2020} found that their $g$-band GCLF can be described by a Gaussian with a mean of $\mu = -7.5\,$mag and a width of $\sigma = 1.0$\,mag, which is consistent with  similar dwarf galaxies in the same galaxy cluster. 

{ We selected the absolute magnitudes of the GCs such that they follow the mentioned Gaussian distribution, written}
as $M_k = \sqrt{2}\sigma \erf^{-1}\left[2\left(\frac{1}{2n}+\frac{k}{n}\right)-1\right]+\mu$ for $k = 0, 1, \ldots, n-1$.
These magnitudes were then converted to luminosities adopting the $g$-band absolute magnitude of the Sun of  5.12 \citep{sparke00}. Finally  the luminosities were converted to masses by multiplying by 2.2, following \citet{spitler09}, who derived this as the typical  value of the mass-to-light ratio of GCs in the $V$ band, a photometric band similar to $g$.

Similarly to the previous section, we plotted in \fig{rtallgclf} the positions of all GCs for every time step for all ten simulations with this more typical GC mass function. The GC system becomes more concentrated {with} time again and we can again note a mass segregation. The evolution of the GC system is nevertheless notably slower than with the DF2-like GC mass function. The profiles of the GC system at different times after the beginning of the simulation are shown in the right column of \fig{gcsprofs}.

{ Figure~\ref{fig:veldisp} shows that the line-of-sight velocity dispersion of the GC system again decreases with time, but the decrease is weaker than in the simulation with the GCs with the DF2-like masses. As \fig{siglosprof} shows, there is nevertheless still a strong drop of the central velocity dispersion of the GC system. The central velocity dispersion of the stars of the galaxy again increased but less prominently than in the simulation with the massive GCs. It can be noted from Figs.~\ref{fig:veldisp} and~\ref{fig:siglosprof} that our simulations predict that the observed line-of-sight velocity dispersion of a GC system of a UDG should be lower than that of the stars of the galaxy. There might be exceptions from this rule caused by the low number of GCs.}

%%%%%%%%%%%%%%%%%%%%%%%%%%%%%%%%%%%%%%%%%%%%%%%%%%%%%%%% 
%%%%%%%%%%%%%%%%%%%%%%%%%%%%%%%%%%%%%%%%%%%%%%%%%%%%%%%%
\section{Extension of the results to other UDGs}
\label{sec:scaling}
Finally, we explore the scaling relations of the evolution of GC systems. This is useful for applying the results of our simulations to galaxies that have properties different from those of the galaxy in our simulations. We did this via a dimensional analysis. We were interested in the scaling of the cumulative fraction of GCs $f$ within a galactocentric radius $r$ at time $t$ after the creation of the GC system. We had to assume that GC systems are formed with properties similar  to the initial conditions of our simulations. Apart from $r$ and $t$, $f$ has to depend on the mass of the galaxy, $M$, its effective radius, $R_\mathrm{e}$, the typical mass of the GC in the GC system, $m$, and on the constants $G$ and $a_0$. For simplicity we assumed that all GCs in the GC system have the same mass, galaxies and the GC systems always have the  S\'ersic indices as in our simulation, the GC systems and galaxies always have a fixed ratio of effective radii, GC systems always contain the same number of GCs, and that the whole system is in the deep-MOND regime. The last assumption implies that in the expression for $f$, $G$ and $a_0$ and all masses $M_i$ can appear only together in the products of the form of $Ga_0M_i$ \citep{milgped, milg08b,milg09b}. Dimensional analysis then implies that $f$ has to have the form of
\begin{equation}
    f = f\left(\frac{m}{M}, \frac{r}{R_\mathrm{e}}, \frac{t\sqrt[4]{GMa_0}}{R_\mathrm{e}}\right).
    \label{eq:scaling}
\end{equation}
The cumulative fraction of GCs within radius $r$ is invariant for all configurations of the problem that keep the arguments of the function on the right-hand side invariant. 

\begin{figure}
        \resizebox{\hsize}{!}{\includegraphics{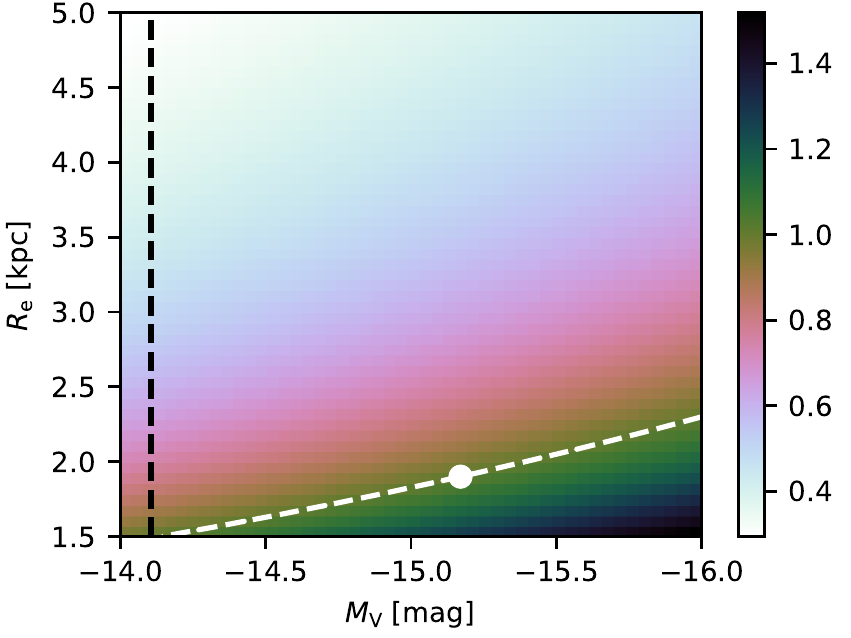}}
        \caption{Time scaling factor given by \equ{tscaling} as a function of the absolute magnitude and effective radius of a galaxy. The white point indicates our simulated galaxy. The white dashed line indicates where the scaling factor equals 1.  {The black dashed line denotes where the GC-mass scaling factor, the second factor in \equ{mscaling}, equals 2.66, which is the value where an observed UDG with GCs following the standard GCLF mass function has to be compared to our simulations with the DF2-like GC mass function.}
        Supposing that the galaxy has a GC system following the standard GC mass function, our results imply that the galaxies to the right of the black line and to the top the white line do not create a nuclear star cluster by inspiralling of GCs.}
        \label{fig:tscaling}
\end{figure}

We now illustrate the use of this formula on an example. Let us assume that we want to compare our simulation to an observed galaxy that has a mass of $M_\mathrm{o} = 0.7\times10^{10}\,M_\sun$, an effective radius of $R_\mathrm{e,o} = 3\,$kpc, a S\'ersic index of one and whose GCs follow the GCLF, such that their average mass is $m_\mathrm{o} = 2.6\times10^{5}\,M_\sun$, and they are $t_\mathrm{o} = 10\,$Gyr old. Let the symbols indexed by ``s'' denote the analogous quantities in the simulation. Equation~\ref{eq:scaling} indicates that we have to compare the observed galaxy to our simulation with the DF2-like GC mass function because the first argument of the function in \equ{scaling} is invariant only if 
\begin{equation}
m_\mathrm{s} = m_\mathrm{o}\frac{M_\mathrm{s}}{M_\mathrm{o}},
\label{eq:mscaling}
\end{equation}
which equals $7.4\times10^{5}\,M_\sun$. The third argument of the function in \equ{scaling} is invariant if 
\begin{equation}
t_\mathrm{s} = t_\mathrm{o}\sqrt[4]{\frac{M_\mathrm{o}}{M_\mathrm{s}}}\frac{R_\mathrm{e,s}}{R_\mathrm{e,o}},
\label{eq:tscaling}
\end{equation}
or 5.8\,Gyr in our case, which is the simulation time to be compared to the observation.

We now applied this scaling relation to investigate the survivability of the GC systems of all UDGs. In \fig{tscaling} we plotted the time scaling factor from \equ{tscaling}, that is, the product of the last two terms, as a function of the absolute $V$-band magnitude of the galaxy and of its effective radius.   We assumed the mass-to-light ratio of the galaxy to be two.  The displayed range of the parameters on the axes of this plot covers most of the observed UDGs.  If a UDG has a standard GCLF mass function, then our simulations, in combination with the scaling relations, show that the GC system does not form a nuclear star cluster if the parameters of the UDG lie above the white dashed line in \fig{tscaling} and to the right of the black dashed line. Nearly all UDGs satisfy this condition; see the data in  \citet{munoz15,vandokkum15,roman17,habas20}. Thus we deduce, that for the majority of UDGs, GC systems do not merge into a nuclear star cluster in MOND.

%%%%%%%%%%%%%%%%%%%%%%%%%%%%%%%%%%%%%%%%%%%%%%%%%%%%%%%%%%%%%%%%%%%%%%%%%%%%%%%
%%%%%%%%%%%%%%%%%%%%%%%%%%%%%%%%%%%%%%%%%%%%%%%%%%%%%%%%%%%%%%%%%%%%%%%%%%%%%%

\section{Influence of the internal structure of the GCs}
\label{sec:inter}
We simulated the GCs by point masses. The real GCs consist of many stars. Therefore, we might miss some important phenomena. For example, GCs can merge or evaporate because of their mutual encounters. The GC system could potentially also sink faster because its orbital energy would be transferred  into the internal energy of GCs during the encounters. Similar questions were already studied for the DF2 galaxy in the Newtonian context by \citet{duttachowdhury20} with simulations and analytically. They found that, in this context, hardly any GC { mergers} can be expected in 10\,Gyr. Similarly, the internal structure of GCs evolved very weakly. In this section, we discuss the situation for GCs of an isolated UDG in MOND.  In this context, because of the creation of the central dense core of the GC system, the interactions between the clusters are stronger and more frequent than in the Newtonian case. It would be ideal to investigate the questions above by simulations with resolved GC. Since this is highly computationally demanding, we progress in this initial study by coupling the many GC simulations described in \sect{manygc} with analytic estimates. 

\subsection{Destruction of GCs by encounters}
We investigated the possibility that the evaporation of {GCs is}  caused by tidal shocks during GC-GC encounters.
We took all pairs of GCs in every simulation and searched for the times of the relative pericenters of the GC pairs. At every pericenter, we determined the relative distance and tangential velocity of the GCs. We took these as the input for calculating the energy gain of each of the GCs through the impulse approximation. The total gain of energy of a particular GC by encounters $\Delta E_\mathrm{enc}$ was obtained by summing over all pericenters with respect to all other GCs in the system during the course of the whole simulation.  We approximated the GCs as Newtonian Plummer spheres because most GCs are high-acceleration objects. 

In particular, the gain of the internal energy of a given GC in one encounter, $\Delta E_\mathrm{1, enc}$, was calculated using the formula for the impulse approximation provided by \citet{banik21}. Unlike the classical formula \citep{spitzer58,mo}, it takes into account the actual density profiles of the interacting bodies. It is therefore suitable even for the close encounters of GCs that occurred in our simulations. The impulse approximation, nevertheless, is designed only for fast encounters, the so-called tidal shocks. This means that we can consider the stars of the investigated clusters stationary with respect to the cluster center during the characteristic duration of the encounter. In our simulation, this is often not satisfied because the velocity dispersion of the GC system (\fig{veldisp}) is not much higher than the velocity dispersion of the stars inside the GCs, which is around 11\,km\,s$^{-1}$ \citep{duttachowdhury20}. When the encounter is very slow { (when there is adiabatic variation of the gravitational potential)}, the energy gain of the cluster after the encounter is zero. This effect can be accounted for by multiplying  the energy gain of a GC predicted by the impulse approximation formula by the ``adiabatic correction'' factor $A = \left[1+(\omega/\tau)^2\right]^{-\gamma}$ \citep{gnedin99}. The typical angular velocity of a star in the cluster $\omega$ was evaluated as  the ratio of the velocity dispersion at the half-mass radius of the cluster, $r_\mathrm{h}$ and $r_\mathrm{h}$ itself. The variable $\tau$ stands for the timescale of the encounter, which was evaluated as the pericentric distance of the encounter divided by the pericentric velocity. Following \citet{banik21}, we put $\gamma = 2-0.5\erf\left[(\tau-2.5t_\mathrm{dyn})/(0.7t_\mathrm{dyn})\right]$, where $t_\mathrm{dyn} = \sqrt{\pi^2r_\mathrm{h}^3/(2GM)}$ is the dynamical time at the half-mass radius of the cluster of the mass $M$. 
In order to decide whether the gain of energy of a GC by the encounters can cause a substantial evaporation of GCs or a creation of tidal tails, we compared $\Delta E_\mathrm{enc}$ to the Newtonian gravitational binding energy of a Plummer sphere, $3\pi GM^2/(32r_\mathrm{h})$. The effective radius of GCs of DF2 is around 8\,pc \citep{duttachowdhury20}, while the MOND transitional radius, where deviations from Newtonian gravity become important, is 34\,pc. If a GC became that big, it would not be classified as a GC, and this is why we consider the Newtonian binding energy. The radii and masses of the GCs of DF2 do not show an obvious correlation and therefore we adopted a universal radius for all GCs. 

For the simulation with the DF2-like GC masses, it is suitable to accept the characteristic radius for the GCs as the typical radius of the GCs of DF2, 8\,pc \citep{duttachowdhury20}. For this value we got a total relative change of internal energy greater than (0.1,0.5,1) for (66,61,57) GCs out of the total 100 in the whole ensemble of our 10 simulations with the DF2-like GC masses. Adopting the characteristic radius of the GCs instead as the maximum scale radius of the GCs of the DF2 galaxy (12\,pc) \citep{duttachowdhury20}, we got a total relative change of internal energy greater than (0.1,0.5,1) for (67,62,58) cases.  This suggests that a substantial fraction of GCs can be affected by evaporation of GCs due to the encounters. Most of the interactions leading to big gains of internal energy happened during the interactions involving the three most massive GCs. For the simulation with the standard GC masses, we first adopted the characteristic radius of a GC in a typical galaxy (3\,pc) \citep{masters10}. For this value we got a total relative change of internal energy greater than (0.1,0.5,1) for (14,4,4) GCs. Adopting instead a twice larger characteristic radius for the GCs (6\,pc), we got a total relative change of internal energy greater than (0.1,0.5,1) in (17,12,6) cases in the whole ensemble of 10 simulations. In other words, for these lower GC masses, the structure of GCs by GC-GC encounters can affect a small fraction of GCs.

\subsection{Mergers between GCs}
In order for two GCs to merge, they first have to become bound. After that, they have to lose { relative} orbital kinetic and potential energy. We therefore estimated the frequency of mergers in two steps { that we describe in detail below}: First, we detected pairs of GCs in the simulations in relative pericenters and decided the GCs would be bound. Second, we checked whether they had dispositions to transfer the orbital energy into the internal energy.

Let us define the function $U(M_1, M_2, b,r)$ as the binding potential energy of two Plummer spheres with masses $M_1$, $M_2$, the characteristic radius $b$ being separated by the distance $r$. The energy $U$ consists of three contributions: the two-body energy $U_\mathrm{2b}$ necessary to separate the two Plummer spheres to infinity and the binding energy of the two Plummer spheres themselves.  Let us further assume that the centers of the Plummer spheres follow the average positions and velocities of the stars that originally belonged to the two interacting GCs. Let $K$ be the relative kinetic energy of the spheres. Before the merger, neglecting the influence of all the other bodies in the system,   the two clusters oscillate in the $U-K$ plane, keeping the sum $E_\mathrm{ini}=U+K$ constant. After the completion of the merger, $K=0$ and $U$ is minimized at $U(r=0)$. In order for the two isolated GCs to merge, the energy $E_\mathrm{merg}=E_\mathrm{ini}-U(r=0)$ has to be transferred { to} the internal energy of the GCs.

For every pair of GCs in a simulation we identified the {instances} of pericenters and calculated the energy loss by tidal shock as above. A GC pair was considered bound if {the difference of the energy of the GC-GC system, $U_\mathrm{2b}+K,$ and of the internal energy of the GCs gained by the shock} was less than zero. In order for a { bound} GC pair to merge, a substantial amount of orbital energy has to be transferred { to the internal energy of} the GCs. If the orbital energy is not lost, encounters with other GCs can make the pair unbound or the pair does not merge during the whole life of the galaxy.  We thus  determined the candidates for merging clusters such that the loss of energy by the current encounter by tidal shock, $\Delta E_\mathrm{1, enc}$, was at least  $0.1E_\mathrm{merg}$, for $U$ and $K$ evaluated at the pericenter. 

In MOND,  the binding energy depends on the value of the external field. For isolated objects, it is infinite because isolated objects produce infinitely deep potential wells in MOND. For our purposes, we assumed that the GC pairs are embedded in an external field of the intensity $0.1a_0$, which is the typical gravitational acceleration imposed by the galaxy itself on the GCs close to the center of the galaxy (\fig{ics}), where we expect most of the interactions between the GCs.  The binding energy of two point masses was obtained by integrating the MOND approximate two-body force formula in the presence of an external field by \citet{zhao13}. Following \citet{bil18}, we included a softening parameter equal to the accepted characteristic radius of the GCs to account for the non-zero size of the GCs. The potential binding energy of a Plummer sphere, necessary for evaluating the energy $U$, was calculated as
\begin{equation}
    U_\mathrm{Plummer} = \int_0^\infty 4\pi G r_1^2\rho(r_1)\int_{r_1}^\infty \mathcal{T}\left[\frac{-GM(<r_1)}{r_2^2}\right]\mathrm{d}r_2\mathrm{d}r_1,
    \label{eq:bind}
\end{equation}
where $\rho$ and $M(<r)$ stand for the  standard density and cumulative mass of a Plummer sphere, respectively, and $\mathcal{T}$ for a function transforming the Newtonian acceleration to the MOND acceleration through the approximate ``1D'' formula for the EFE in QUMOND \citep{famaey12}. The post-merger energy $U(r=0)$ was evaluated from \equ{bind}, when considering the two Plummer spheres overlaid { on} each other.

In this way, we obtained for the simulation with the DF2-like GC masses  15 candidate merging GC pairs in the whole set of 10 simulations for the characteristic radius of the GCs of 8\,pc. Adopting the maximum GC characteristic radius in DF2, 12\,pc, we found 8  GC merger candidates pairs. This suggests that for isolated UDGs with GCs similar to DF2, GC mergers can occasionally happen. For the standard GC masses, we obtained no mergers regardless of whether we assumed the GC radii of 3 or 6 pc.

\subsection{Loss of orbital energy of GCs from GC interactions}
We also explored the possibility that the GCs might sink toward the center of the UDG  faster because their orbital energy with respect to the galaxy would be transferred to the internal energy of the GCs because of their interactions with each other. We therefore again detected the times of pericenters of all GC pairs in the simulations and calculated the energy absorbed by the GCs  via the tidal approximation approach described above. We then compared the sum of all of these energy losses to the  total kinetic energy of all GCs at the end of the simulation. 

For the DF2-like GC masses and a characteristic radius of 8\,pc, we found that the energy loss is 2.8 times higher than the kinetic energy at the end of the simulation.  Adopting the GC radius of 12\,pc, 2.3 times the final kinetic energy of GCs is transferred to the internal energy of GCs. This indicates that the GC-GC interactions might also have an effect on the spatial structure of the GC system. This is not the case for the standard GC masses. If their characteristic radius is assumed to be 3\,pc, the absorbed energy makes only 0.05 of the kinetic energy of the GC system at the end of the simulation. Adopting the double GC radius, the GC system transferred a fraction of 0.07 of its kinetic energy to the internal energy of the GCs.

%%%%%%%%%%%%%%%%%%%%%%%%%%%%%%%%%%%%%%%%%%%%%%%
%%%%%%%%%%%%%%%%%%%%%%%%%%%%%%%%%%%%%%%%%%%%%%%
\section{Summary and conclusions}
\label{sec:conclusions}
Dynamical friction in MOND gravity is far less explored than in Newtonian gravity. In this contribution, we aimed to improve this deficiency. We first sought to test for the first time the validity of the proposed MOND analog of the \cf using high-resolution, self-consistent simulations. We started by studying dynamical friction in an example of a single GC, represented by a point mass, moving in the gravitational field of a UDG with a mass of $2 \times 10^8 M_\sun$ similar to the DF2 galaxy (but in isolation). We found that the \ssf (\equ{ssf}) works excellently for a suitable choice of the effective value of  the Coulomb logarithm, as long as the GC does not have its apocenter closer than about 0.75\,kpc to the center of the UDG. For radial orbits and realistic masses of the GC (i.e., less than about $2\times10^6\,M_\sun$), the best value of the Coulomb logarithm is around $\ln\Lambda \simeq 3$. For higher masses of the GC, between $2\times10^6$ and $2\times10^7\,M_\sun$ (one-tenth of the mass of the galaxy), the value of the Coulomb logarithm decreases. This is in line with the findings of  previous works that dynamical friction in MOND becomes ineffective for interactions of objects with comparable masses. On the other hand, the effective value of the Coulomb logarithm is higher for non-radial orbits. This  agrees with the results of Newtonian simulations. It can likely be explained by a higher contribution  of the resonant coupling mechanism to dynamical friction.  The effective value of the Coulomb logarithm depends not only on the circularity of the orbit, but also on its size: the highest values were encountered for orbits close to the center of the galaxy. The highest effective value of the Coulomb logarithm that we encountered in our simulations was $\ln \Lambda = 15$.

The \ssf stopped being applicable to our problem  when the apocentric distance of the GC decreased below about 0.75\,kpc ($0.4\,R_\mathrm{e}$) with respect to the center of the galaxy. Dynamical friction becomes much less effective in this central region. The GCs in our simulations, even the most massive ones, never sink  right to the center of the galaxy, even if predicted to be the case by the \ssf. We verified that this is not because of insufficient resolution of the simulations.  A similar effect, known from Newtonian simulations, is known as core stalling. Whether core stalling occurs in Newtonian simulations depends on the inner density profile of the galaxy, but we did not explore this aspect in MOND. For a GC of $10^6\,M_\sun$ falling from 5\,kpc ($2.5\,R_\mathrm{e}$), it takes about 3\,Gyr to reach the stalling phase, but the same takes more than 10\,Gyr for a GC with a mass of $10^5\,M_\sun$. These inspiralling times depends on the particular density distribution of the galaxy. { In total, we have shown that the \ssf can serve as a rule of thumb to estimate the order of dynamical friction.  It has a limited range of applicability, but in certain circumstances it is a very good approximation.}

We also explored the evolution of a whole GC system consisting of ten GCs around the UDG. We considered two types of GC systems: the GCs were either ``overmassive'' as deduced in DF2 (if it is at a distance of 20\,Mpc), or the GCs had masses following the standard GC mass function. 
It is not established yet how often anomalous GC mass functions occur in UDGs. Follow-up observations of another UDG with a suspected DF2-like mass function \citep{2020A&A...640A.106M} have shown that it rather follows a standard GCLF mass function \citep{2021arXiv210110659M}.
 
The more massive GC system undertakes a dramatic structural change during the first 6\,Gyr of its evolution, when its half-number radius changes from 3.5\,kpc to about 0.6\,kpc.  The most massive GCs sink more quickly. After that time, the half-number radius remains nearly constant up to the end of the simulation at 10\,Gyr, at which point most of the GCs have reached their core-stalling phase. Even in this phase, the GCs would likely be too far from each other to merge together to form a nuclear star cluster, since the average radius of { the massive GCs of UDGs}  is only 6-8\,pc \citep{vandokkum18}. Hence we need to be careful when arguing against MOND on the basis of the survival of the GC system with analytic formulas such as \equ{ssf}. 
{ We note that we  simulated a galaxy that is in isolation. While its parameters were inspired by the DF2 galaxy, the real DF2 galaxy, if indeed at the measured distance, has to be close to its neighboring galaxy NGC\,1052 dominated by its external field and therefore its internal dynamics is effectively Newtonian. The GC system of DF2 in Newtonian dynamics has been studied by \citet{duttachowdhury19}, who found that it is consistent with observations if it was somewhat more extended in the past. The GC system of DF2 is thus consistent with Newtonian and MOND gravity. The fact that the GC system of DF2 does not have a core (see \sect{df2mf}) then suggests that DF2 has spend most of its life dominated by the external field of NGC\,1052.}

The { simulated} GC system following the standard GC mass function experiences a weaker evolution, decreasing its 3D half-number radius from 3.5\,kpc to about 2\,kpc in 10\,Gyr. Extrapolating this result backward in time, we expect that GC systems of isolated UDGs were larger by a factor of almost 2 at their formation 10\,Gyr ago with the standard GCLF.   

For both choices of the mass function of the GC system, we detected a mass segregation of the GCs. This can be taken as a prediction for observations. We found that the GC system of the real DF2 galaxy has no signature of the central core. This indicates that in MOND, if DF2 is a member of the NGC\,1052 group, it has to have spent a large fraction of its life dominated by the EFE imposed by this galaxy. Further, we were able to show by scaling our simulations for most GC systems of UDGs that they are not expected to merge and form nuclear star clusters.

{ Not only the spatial distribution of GCs evolved in our simulations, but also the central velocity dispersion of the GC system decreased and that of the stars of the galaxy increased, likely because of energy equipartition (see Figs.~\ref{fig:veldisp} and~\ref{fig:siglosprof}). We can thus expect that in observations, UDGs usually have a higher velocity dispersion of stars than of GC systems. 
}

{ Our simulations were limited in the sense that the GCs were modeled as point masses instead of bodies consisting of many particles. The many GC simulations were thus not able to capture the effects related to absorbing the orbital energy of the GC system in the internal energy of the individual GCs. We partly resolved this issue, at least for a UDG of the properties of the simulated galaxy, by joining the simulations with analytic estimates. For the simulations with a GC system inspired by DF2, we found that GC-GC interactions are expected to play a substantial role in the evolution of the GC system.  About one-half of the GCs are expected to receive energy that is comparable to their internal Newtonian binding energy. Therefore, not only the GCs might lose a large fraction of their stars because of the encounters, but the GC system might sink substantially faster to the center of the galaxy because of the loss of orbital energy. For the DF2-like GCs, we therefore still cannot reliably exclude the possibility that some of the GCs would form a nuclear star cluster in an isolated UDG. This is to be investigated by dedicated simulations with GCs simulated as many-body systems. For the standard GCs, which seem to be the most common even in UDGs, our analytic estimates indicate a much less dramatic influence of mutual GC interactions. The increase in internal energy of GCs is expected to be comparable to the Newtonian binding energy of GCs for only about one GC out of the ten in the GC system. We estimated that the GC-GC mergers are nearly non-existent, and there is no acceleration of the sinking of the GC system to the center of the UDG. We thus do not expect that the GCs would form a nuclear star cluster if the GCs were simulated as many-particle systems.}

The question of survivability of GC systems in MOND  remains open for dwarf galaxies. Such galaxies have lower effective radii and the same or lower brightnesses than UDGs. For example, for a dwarf galaxy similar to the Fornax dwarf, having a stellar mass of $3\times10^7\,M_\sun$ and an effective radius of 0.5\,kpc, we deduce a GC-mass scaling factor of 6.7 (\equ{mscaling}) and a time scaling factor of 2.6 (\equ{tscaling}). The evolution of the GC system due to dynamical friction is thus much faster than in our simulation of the UDG with the DF2-like GC mass function.  It is difficult to predict from our present simulations whether a nuclear star cluster would form. Observations of chemical properties of some dwarf galaxies suggest that their nuclear star clusters formed by dynamical sinking of GCs \citep{fahrion20}; the galaxies had much smaller radii and brightnesses than the UDGs investigated in this work. Future work should address in which galaxies MOND predicts the formation of a nuclear star cluster through the merging of the GC system.

Previous simulations showed that major mergers are much less effective in MOND than in Newtonian simulations with dark matter. Analytic arguments however indicated that the situation is the opposite for minor mergers around spheroidal galaxies. Our simulations confirmed, for the first time, the analytic prediction for situations similar to small galaxies being accreted by much more massive spheroidal galaxies (previous works confirmed them only for the rotation of galactic bars). This forces us to consider seriously the potential importance of minor mergers for galaxy evolution in MOND. Nevertheless, minor merging will require a more careful examination. The first reason is that Newtonian simulations show that the size of the satellite can also influence dynamical friction \citep{mo}, while our simulated  satellite was a point mass. It should also be explored better how dynamical friction works in MOND when the satellite moves outside of the main galaxy, where the density of stars is negligible. In Newtonian gravity, in these regions we still have dark matter halos up to hundreds of kiloparsec from the galaxy that can absorb the energy of the satellite. These are questions are to be investigated by future works.

\begin{acknowledgements}
{ The authors thank the anonymous referee for a constructive report which helped to significantly improve the paper.}
MB and HSZ are thankful for the financial support by {\it Cercle Gutenberg}. MB acknowledges the support from the Polish National Science Centre under the grant 2017/26/D/ST9/00449. BF and RI acknowledge financial support from the European Research Council (ERC) under the European Unions Horizon 2020 research and innovation programme (grant agreement No. 834148). 
 O.M. is grateful to the Swiss National Science Foundation for financial support. 
\end{acknowledgements}

\bibliographystyle{aa}
\bibliography{literature}

\begin{thebibliography}{113}
\expandafter\ifx\csname natexlab\endcsname\relax\def\natexlab#1{#1}\fi

\bibitem[{{Angus} \& {Diaferio}(2009)}]{angus09}
{Angus}, G.~W. \& {Diaferio}, A. 2009, \mnras, 396, 887

\bibitem[{{Asencio} {et~al.}(2021){Asencio}, {Banik}, \& {Kroupa}}]{asencio21}
{Asencio}, E., {Banik}, I., \& {Kroupa}, P. 2021, \mnras, 500, 5249

\bibitem[{{Banik} \& {Zhao}(2018)}]{banik18b}
{Banik}, I. \& {Zhao}, H. 2018, \mnras, 480, 2660

\bibitem[{{Banik} \& {van den Bosch}(2021)}]{banik21}
{Banik}, U. \& {van den Bosch}, F.~C. 2021, \mnras, 502, 1441

\bibitem[{{Beasley} {et~al.}(2016){Beasley}, {Romanowsky}, {Pota}, {Navarro},
  {Martinez Delgado}, {Neyer}, \& {Deich}}]{beasley16}
{Beasley}, M.~A., {Romanowsky}, A.~J., {Pota}, V., {et~al.} 2016, \apjl, 819,
  L20

\bibitem[{{Bekenstein} \& {Milgrom}(1984)}]{bm84}
{Bekenstein}, J. \& {Milgrom}, M. 1984, \apj, 286, 7

\bibitem[{{B{\'{\i}}lek} {et~al.}(2019){B{\'{\i}}lek}, {Samurovi{\'c}}, \&
  {Renaud}}]{bil19}
{B{\'{\i}}lek}, M., {Samurovi{\'c}}, S., \& {Renaud}, F. 2019, \aap, 625, A32

\bibitem[{{B{\'{\i}}lek} {et~al.}(2018){B{\'{\i}}lek}, {Thies}, {Kroupa}, \&
  {Famaey}}]{bil18}
{B{\'{\i}}lek}, M., {Thies}, I., {Kroupa}, P., \& {Famaey}, B. 2018, \aap, 614,
  A59

\bibitem[{{Caldwell} {et~al.}(2017){Caldwell}, {Walker}, {Mateo}, {Olszewski},
  {Koposov}, {Belokurov}, {Torrealba}, {Geringer-Sameth}, \&
  {Johnson}}]{caldwell17}
{Caldwell}, N., {Walker}, M.~G., {Mateo}, M., {et~al.} 2017, \apj, 839, 20

\bibitem[{{Carleton} {et~al.}(2019){Carleton}, {Errani}, {Cooper},
  {Kaplinghat}, {Pe{\~n}arrubia}, \& {Guo}}]{Carleton2019}
{Carleton}, T., {Errani}, R., {Cooper}, M., {et~al.} 2019, \mnras, 485, 382

\bibitem[{{Chae} {et~al.}(2020){Chae}, {Lelli}, {Desmond}, {McGaugh}, {Li}, \&
  {Schombert}}]{chae20}
{Chae}, K.-H., {Lelli}, F., {Desmond}, H., {et~al.} 2020, \apj, 904, 51

\bibitem[{{Chan} {et~al.}(1997){Chan}, {Mamon}, \& {Gerbal}}]{chan97}
{Chan}, R., {Mamon}, G.~A., \& {Gerbal}, D. 1997, Astrophysical Letters and
  Communications, 36, 47

\bibitem[{{Chandrasekhar}(1943)}]{chandrasekhar43}
{Chandrasekhar}, S. 1943, \apj, 97, 255

\bibitem[{{Ciotti} \& {Binney}(2004)}]{ciotti04}
{Ciotti}, L. \& {Binney}, J. 2004, \mnras, 351, 285

\bibitem[{{Combes}(2014)}]{combes14}
{Combes}, F. 2014, \aap, 571, A82

\bibitem[{{Crnojevi{\'c}} {et~al.}(2014){Crnojevi{\'c}}, {Sand}, {Caldwell},
  {Guhathakurta}, {McLeod}, {Seth}, {Simon}, {Strader}, \&
  {Toloba}}]{crnojevic14}
{Crnojevi{\'c}}, D., {Sand}, D.~J., {Caldwell}, N., {et~al.} 2014, \apjl, 795,
  L35

\bibitem[{{Danieli} {et~al.}(2020){Danieli}, {van Dokkum}, {Abraham}, {Conroy},
  {Dolphin}, \& {Romanowsky}}]{danieli20}
{Danieli}, S., {van Dokkum}, P., {Abraham}, R., {et~al.} 2020, \apjl, 895, L4

\bibitem[{{Durrell} {et~al.}(1996){Durrell}, {Harris}, {Geisler}, \&
  {Pudritz}}]{1996AJ....112..972D}
{Durrell}, P.~R., {Harris}, W.~E., {Geisler}, D., \& {Pudritz}, R.~E. 1996,
  \aj, 112, 972

\bibitem[{{Dutta Chowdhury} {et~al.}(2019){Dutta Chowdhury}, {van den Bosch},
  \& {van Dokkum}}]{duttachowdhury19}
{Dutta Chowdhury}, D., {van den Bosch}, F.~C., \& {van Dokkum}, P. 2019, \apj,
  877, 133

\bibitem[{{Dutta Chowdhury} {et~al.}(2020){Dutta Chowdhury}, {van den Bosch},
  \& {van Dokkum}}]{duttachowdhury20}
{Dutta Chowdhury}, D., {van den Bosch}, F.~C., \& {van Dokkum}, P. 2020, \apj,
  903, 149

\bibitem[{{Fahrion} {et~al.}(2020){Fahrion}, {M{\"u}ller}, {Rejkuba}, {Hilker},
  {Lyubenova}, {van de Ven}, {Georgiev}, {Lelli}, {Pawlowski}, \&
  {Jerjen}}]{fahrion20}
{Fahrion}, K., {M{\"u}ller}, O., {Rejkuba}, M., {et~al.} 2020, \aap, 634, A53

\bibitem[{{Famaey} {et~al.}(2018){Famaey}, {McGaugh}, \& {Milgrom}}]{famaey18}
{Famaey}, B., {McGaugh}, S., \& {Milgrom}, M. 2018, \mnras, 480, 473

\bibitem[{{Famaey} \& {McGaugh}(2012)}]{famaey12}
{Famaey}, B. \& {McGaugh}, S.~S. 2012, Living Reviews in Relativity, 15, 10

\bibitem[{{Freundlich} {et~al.}(2020){Freundlich}, {Dekel}, {Jiang}, {Ishai},
  {Cornuault}, {Lapiner}, {Dutton}, \& {Macci{\`o}}}]{Freundlich20}
{Freundlich}, J., {Dekel}, A., {Jiang}, F., {et~al.} 2020, \mnras, 491, 4523

\bibitem[{{Georgiev} {et~al.}(2009){Georgiev}, {Puzia}, {Hilker}, \&
  {Goudfrooij}}]{2009MNRAS.392..879G}
{Georgiev}, I.~Y., {Puzia}, T.~H., {Hilker}, M., \& {Goudfrooij}, P. 2009,
  \mnras, 392, 879

\bibitem[{{Gnedin} \& {Ostriker}(1999)}]{gnedin99}
{Gnedin}, O.~Y. \& {Ostriker}, J.~P. 1999, \apj, 513, 626

\bibitem[{{Greco} {et~al.}(2018){Greco}, {Greene}, {Strauss}, {Macarthur},
  {Flowers}, {Goulding}, {Huang}, {Kim}, {Komiyama}, {Leauthaud}, {Leisman},
  {Lupton}, {Sif{\'o}n}, \& {Wang}}]{Greco2018}
{Greco}, J.~P., {Greene}, J.~E., {Strauss}, M.~A., {et~al.} 2018, \apj, 857,
  104

\bibitem[{{Habas} {et~al.}(2020{\natexlab{a}}){Habas}, {Marleau}, {Duc},
  {Durrell}, {Paudel}, {Poulain}, {S{\'a}nchez-Janssen}, {Sreejith},
  {Ramasawmy}, {Stemock}, {Leach}, {Cuillandre}, {Gwyn}, {Agnello},
  {B{\'\i}lek}, {Fensch}, {M{\"u}ller}, {Peng}, \& {van der Burg}}]{habas20}
{Habas}, R., {Marleau}, F.~R., {Duc}, P.-A., {et~al.} 2020{\natexlab{a}},
  \mnras, 491, 1901

\bibitem[{{Habas} {et~al.}(2020{\natexlab{b}}){Habas}, {Marleau}, {Duc},
  {Durrell}, {Paudel}, {Poulain}, {S{\'a}nchez-Janssen}, {Sreejith},
  {Ramasawmy}, {Stemock}, {Leach}, {Cuillandre}, {Gwyn}, {Agnello},
  {B{\'\i}lek}, {Fensch}, {M{\"u}ller}, {Peng}, \& {van der
  Burg}}]{2020MNRAS.491.1901H}
{Habas}, R., {Marleau}, F.~R., {Duc}, P.-A., {et~al.} 2020{\natexlab{b}},
  \mnras, 491, 1901

\bibitem[{{Haghi} {et~al.}(2016){Haghi}, {Bazkiaei}, {Zonoozi}, \&
  {Kroupa}}]{haghi16}
{Haghi}, H., {Bazkiaei}, A.~E., {Zonoozi}, A.~H., \& {Kroupa}, P. 2016, \mnras,
  458, 4172

\bibitem[{{Haghi} {et~al.}(2019){Haghi}, {Kroupa}, {Banik}, {Wu}, {Zonoozi},
  {Javanmardi}, {Ghari}, {M{\"u}ller}, {Dabringhausen}, \& {Zhao}}]{haghi19}
{Haghi}, H., {Kroupa}, P., {Banik}, I., {et~al.} 2019, \mnras, 487, 2441

\bibitem[{{Hamilton} {et~al.}(2018){Hamilton}, {Fouvry}, {Binney}, \&
  {Pichon}}]{Hamilton18}
{Hamilton}, C., {Fouvry}, J.-B., {Binney}, J., \& {Pichon}, C. 2018, \mnras,
  481, 2041

\bibitem[{{Haslbauer} {et~al.}(2020){Haslbauer}, {Banik}, \&
  {Kroupa}}]{haslbauer20}
{Haslbauer}, M., {Banik}, I., \& {Kroupa}, P. 2020, \mnras, 499, 2845

\bibitem[{{Hees} {et~al.}(2016){Hees}, {Famaey}, {Angus}, \&
  {Gentile}}]{hees16}
{Hees}, A., {Famaey}, B., {Angus}, G.~W., \& {Gentile}, G. 2016, \mnras, 455,
  449

\bibitem[{{Hernandez} \& {Gilmore}(1998)}]{hernandez98}
{Hernandez}, X. \& {Gilmore}, G. 1998, \mnras, 297, 517

\bibitem[{{Impey} {et~al.}(1988){Impey}, {Bothun}, \& {Malin}}]{impey88}
{Impey}, C., {Bothun}, G., \& {Malin}, D. 1988, \apj, 330, 634

\bibitem[{{Iodice} {et~al.}(2020){Iodice}, {Cantiello}, {Hilker}, {Rejkuba},
  {Arnaboldi}, {Spavone}, {Greggio}, {Forbes}, {D'Ago}, {Mieske}, {Spiniello},
  {La Marca}, {Rampazzo}, {Paolillo}, {Capaccioli}, \& {Schipani}}]{iodice20}
{Iodice}, E., {Cantiello}, M., {Hilker}, M., {et~al.} 2020, \aap, 642, A48

\bibitem[{{Jiang} {et~al.}(2008){Jiang}, {Jing}, {Faltenbacher}, {Lin}, \&
  {Li}}]{jiang08}
{Jiang}, C.~Y., {Jing}, Y.~P., {Faltenbacher}, A., {Lin}, W.~P., \& {Li}, C.
  2008, \apj, 675, 1095

\bibitem[{{Jiang} {et~al.}(2014){Jiang}, {Jing}, \& {Han}}]{jiang14}
{Jiang}, C.~Y., {Jing}, Y.~P., \& {Han}, J. 2014, \apj, 790, 7

\bibitem[{{Jiang} {et~al.}(2019){Jiang}, {Dekel}, {Freundlich}, {Romanowsky},
  {Dutton}, {Macci{\`o}}, \& {Di Cintio}}]{Jiang2019}
{Jiang}, F., {Dekel}, A., {Freundlich}, J., {et~al.} 2019, \mnras, 487, 5272

\bibitem[{{Jord{\'a}n} {et~al.}(2007){Jord{\'a}n}, {McLaughlin},
  {C{\^o}t{\'e}}, {Ferrarese}, {Peng}, {Mei}, {Villegas}, {Merritt}, {Tonry},
  \& {West}}]{jordan07}
{Jord{\'a}n}, A., {McLaughlin}, D.~E., {C{\^o}t{\'e}}, P., {et~al.} 2007,
  \apjs, 171, 101

\bibitem[{{Just} \& {Pe{\~n}arrubia}(2005)}]{just05}
{Just}, A. \& {Pe{\~n}arrubia}, J. 2005, \aap, 431, 861

\bibitem[{{Koda} {et~al.}(2015){Koda}, {Yagi}, {Yamanoi}, \&
  {Komiyama}}]{2015ApJ...807L...2K}
{Koda}, J., {Yagi}, M., {Yamanoi}, H., \& {Komiyama}, Y. 2015, \apjl, 807, L2

\bibitem[{{Kroupa} {et~al.}(2018){Kroupa}, {Haghi}, {Javanmardi}, {Zonoozi},
  {M{\"u}ller}, {Banik}, {Wu}, {Zhao}, \& {Dabringhausen}}]{kroupa18}
{Kroupa}, P., {Haghi}, H., {Javanmardi}, B., {et~al.} 2018, \nat, 561, E4

\bibitem[{{Lacey} \& {Cole}(1993)}]{lacey93}
{Lacey}, C. \& {Cole}, S. 1993, \mnras, 262, 627

\bibitem[{{Leisman} {et~al.}(2017){Leisman}, {Haynes}, {Janowiecki},
  {Hallenbeck}, {J{\'o}zsa}, {Giovanelli}, {Adams}, {Bernal Neira}, {Cannon},
  {Janesh}, {Rhode}, \& {Salzer}}]{leisman17}
{Leisman}, L., {Haynes}, M.~P., {Janowiecki}, S., {et~al.} 2017, \apj, 842, 133

\bibitem[{{Lelli} {et~al.}(2016){Lelli}, {McGaugh}, {Schombert}, \&
  {Pawlowski}}]{lelli16}
{Lelli}, F., {McGaugh}, S.~S., {Schombert}, J.~M., \& {Pawlowski}, M.~S. 2016,
  \apjl, 827, L19

\bibitem[{{Lelli} {et~al.}(2017){Lelli}, {McGaugh}, {Schombert}, \&
  {Pawlowski}}]{lelli17}
{Lelli}, F., {McGaugh}, S.~S., {Schombert}, J.~M., \& {Pawlowski}, M.~S. 2017,
  \apj, 836, 152

\bibitem[{{Lim} {et~al.}(2020){Lim}, {C{\^o}t{\'e}}, {Peng}, {Ferrarese},
  {Roediger}, {Durrell}, {Mihos}, {Wang}, {Gwyn}, {Cuillandre}, {Liu},
  {S{\'a}nchez-Janssen}, {Toloba}, {Sales}, {Guhathakurta}, {Lan{\c{c}}on}, \&
  {Puzia}}]{lim2020}
{Lim}, S., {C{\^o}t{\'e}}, P., {Peng}, E.~W., {et~al.} 2020, \apj, 899, 69

\bibitem[{{Lim} {et~al.}(2018){Lim}, {Peng}, {C{\^o}t{\'e}}, {Sales}, {den
  Brok}, {Blakeslee}, \& {Guhathakurta}}]{2018ApJ...862...82L}
{Lim}, S., {Peng}, E.~W., {C{\^o}t{\'e}}, P., {et~al.} 2018, \apj, 862, 82

\bibitem[{{Lima Neto} {et~al.}(1999){Lima Neto}, {Gerbal}, \&
  {M{\'a}rquez}}]{sersdeproj}
{Lima Neto}, G.~B., {Gerbal}, D., \& {M{\'a}rquez}, I. 1999, \mnras, 309, 481

\bibitem[{{Lin} \& {Tremaine}(1983)}]{lin83}
{Lin}, D.~N.~C. \& {Tremaine}, S. 1983, \apj, 264, 364

\bibitem[{{L{\"u}ghausen} {et~al.}(2015){L{\"u}ghausen}, {Famaey}, \&
  {Kroupa}}]{por}
{L{\"u}ghausen}, F., {Famaey}, B., \& {Kroupa}, P. 2015, Canadian Journal of
  Physics, 93, 232

\bibitem[{{M{\'a}rquez} {et~al.}(2000){M{\'a}rquez}, {Lima Neto}, {Capelato},
  {Durret}, \& {Gerbal}}]{sersdeprojupdate}
{M{\'a}rquez}, I., {Lima Neto}, G.~B., {Capelato}, H., {Durret}, F., \&
  {Gerbal}, D. 2000, \aap, 353, 873

\bibitem[{{Masters} {et~al.}(2010){Masters}, {Jord{\'a}n}, {C{\^o}t{\'e}},
  {Ferrarese}, {Blakeslee}, {Infante}, {Peng}, {Mei}, \& {West}}]{masters10}
{Masters}, K.~L., {Jord{\'a}n}, A., {C{\^o}t{\'e}}, P., {et~al.} 2010, \apj,
  715, 1419

\bibitem[{{McGaugh} \& {Milgrom}(2013{\natexlab{a}})}]{mcgaugh13a}
{McGaugh}, S. \& {Milgrom}, M. 2013{\natexlab{a}}, \apj, 766, 22

\bibitem[{{McGaugh} \& {Milgrom}(2013{\natexlab{b}})}]{mcgaugh13b}
{McGaugh}, S. \& {Milgrom}, M. 2013{\natexlab{b}}, \apj, 775, 139

\bibitem[{{McGaugh}(2016)}]{mcgaugh16b}
{McGaugh}, S.~S. 2016, \apjl, 832, L8

\bibitem[{{McGaugh} {et~al.}(2016){McGaugh}, {Lelli}, \&
  {Schombert}}]{mcgaugh16}
{McGaugh}, S.~S., {Lelli}, F., \& {Schombert}, J.~M. 2016, Physical Review
  Letters, 117, 201101

\bibitem[{{Mihos} {et~al.}(2015){Mihos}, {Durrell}, {Ferrarese}, {Feldmeier},
  {C{\^o}t{\'e}}, {Peng}, {Harding}, {Liu}, {Gwyn}, \& {Cuillandre}}]{mihos15}
{Mihos}, J.~C., {Durrell}, P.~R., {Ferrarese}, L., {et~al.} 2015, \apjl, 809,
  L21

\bibitem[{{Milgrom}(1983{\natexlab{a}})}]{milg83b}
{Milgrom}, M. 1983{\natexlab{a}}, \apj, 270, 371

\bibitem[{{Milgrom}(1983{\natexlab{b}})}]{milg83a}
{Milgrom}, M. 1983{\natexlab{b}}, \apj, 270, 365

\bibitem[{{Milgrom}(1999)}]{milg99}
{Milgrom}, M. 1999, Physics Letters A, 253, 273

\bibitem[{{Milgrom}(2001)}]{milgped}
{Milgrom}, M. 2001, Acta Physica Polonica B, 32, 3613

\bibitem[{{Milgrom}(2008)}]{milg08b}
{Milgrom}, M. 2008, arXiv e-prints, arXiv:0801.3133

\bibitem[{{Milgrom}(2009)}]{milg09b}
{Milgrom}, M. 2009, \apj, 698, 1630

\bibitem[{{Milgrom}(2010)}]{qumond}
{Milgrom}, M. 2010, \mnras, 403, 886

\bibitem[{{Milgrom}(2012)}]{milg12}
{Milgrom}, M. 2012, Physical Review Letters, 109, 131101

\bibitem[{{Milgrom}(2013)}]{milg13}
{Milgrom}, M. 2013, \prl, 111, 041105

\bibitem[{{Mo} {et~al.}(2010){Mo}, {van den Bosch}, \& {White}}]{mo}
{Mo}, H., {van den Bosch}, F.~C., \& {White}, S. 2010, {Galaxy Formation and
  Evolution} (Cambridge Univ. Press, Cambridge)

\bibitem[{{Mu{\~n}oz} {et~al.}(2015){Mu{\~n}oz}, {Eigenthaler}, {Puzia},
  {Taylor}, {Ordenes-Brice{\~n}o}, {Alamo-Mart{\'\i}nez}, {Ribbeck},
  {{\'A}ngel}, {Capaccioli}, {C{\^o}t{\'e}}, {Ferrarese}, {Galaz}, {Hempel},
  {Hilker}, {Jord{\'a}n}, {Lan{\c{c}}on}, {Mieske}, {Paolillo}, {Richtler},
  {S{\'a}nchez-Janssen}, \& {Zhang}}]{munoz15}
{Mu{\~n}oz}, R.~P., {Eigenthaler}, P., {Puzia}, T.~H., {et~al.} 2015, \apjl,
  813, L15

\bibitem[{{M{\"u}ller} {et~al.}(2021{\natexlab{a}}){M{\"u}ller}, {Durrell},
  {Marleau}, {Duc}, {Lim}, {Posti}, {Agnello}, {S{\'a}nchez-Janssen},
  {Poulain}, {Habas}, {Emsellem}, {Paudel}, {van der Burg}, \&
  {Fensch}}]{2021arXiv210110659M}
{M{\"u}ller}, O., {Durrell}, P.~R., {Marleau}, F.~R., {et~al.}
  2021{\natexlab{a}}, arXiv e-prints, arXiv:2101.10659

\bibitem[{{M{\"u}ller} {et~al.}(2019){M{\"u}ller}, {Famaey}, \&
  {Zhao}}]{2019A&A...623A..36M}
{M{\"u}ller}, O., {Famaey}, B., \& {Zhao}, H. 2019, \aap, 623, A36

\bibitem[{{M{\"u}ller} {et~al.}(2018){M{\"u}ller}, {Jerjen}, \&
  {Binggeli}}]{mueller18}
{M{\"u}ller}, O., {Jerjen}, H., \& {Binggeli}, B. 2018, \aap, 615, A105

\bibitem[{{M{\"u}ller} {et~al.}(2020){M{\"u}ller}, {Marleau}, {Duc}, {Habas},
  {Fensch}, {Emsellem}, {Poulain}, {Lim}, {Agnello}, {Durrell}, {Paudel},
  {S{\'a}nchez-Janssen}, \& {van der Burg}}]{2020A&A...640A.106M}
{M{\"u}ller}, O., {Marleau}, F.~R., {Duc}, P.-A., {et~al.} 2020, \aap, 640,
  A106

\bibitem[{{M{\"u}ller} {et~al.}(2021{\natexlab{b}}){M{\"u}ller}, {Pawlowski},
  {Lelli}, {Fahrion}, {Rejkuba}, {Hilker}, {Kanehisa}, {Libeskind}, \&
  {Jerjen}}]{mueller20planes}
{M{\"u}ller}, O., {Pawlowski}, M.~S., {Lelli}, F., {et~al.} 2021{\natexlab{b}},
  \aap, 645, L5

\bibitem[{{Nipoti} {et~al.}(2008){Nipoti}, {Ciotti}, {Binney}, \&
  {Londrillo}}]{nipoti08}
{Nipoti}, C., {Ciotti}, L., {Binney}, J., \& {Londrillo}, P. 2008, \mnras, 386,
  2194

\bibitem[{{Nipoti} {et~al.}(2007){Nipoti}, {Londrillo}, \& {Ciotti}}]{nipoti07}
{Nipoti}, C., {Londrillo}, P., \& {Ciotti}, L. 2007, \mnras, 381, L104

\bibitem[{{Pawlowski}(2018)}]{pawlowski18}
{Pawlowski}, M.~S. 2018, Modern Physics Letters A, 33, 1830004

\bibitem[{{Pawlowski} \& {Kroupa}(2020)}]{pawlowski20}
{Pawlowski}, M.~S. \& {Kroupa}, P. 2020, \mnras, 491, 3042

\bibitem[{{Petts} {et~al.}(2015){Petts}, {Gualandris}, \& {Read}}]{petts15}
{Petts}, J.~A., {Gualandris}, A., \& {Read}, J.~I. 2015, \mnras, 454, 3778

\bibitem[{{Prole} {et~al.}(2019){Prole}, {Hilker}, {van der Burg}, {Cantiello},
  {Venhola}, {Iodice}, {van de Ven}, {Wittmann}, {Peletier}, {Mieske},
  {Capaccioli}, {Napolitano}, {Paolillo}, {Spavone}, \&
  {Valentijn}}]{2019MNRAS.484.4865P}
{Prole}, D.~J., {Hilker}, M., {van der Burg}, R.~F.~J., {et~al.} 2019, \mnras,
  484, 4865

\bibitem[{{Read} {et~al.}(2006){Read}, {Goerdt}, {Moore}, {Pontzen}, {Stadel},
  \& {Lake}}]{read06}
{Read}, J.~I., {Goerdt}, T., {Moore}, B., {et~al.} 2006, \mnras, 373, 1451

\bibitem[{{Rejkuba}(2012)}]{2012Ap&SS.341..195R}
{Rejkuba}, M. 2012, \apss, 341, 195

\bibitem[{{Renaud} {et~al.}(2017){Renaud}, {Agertz}, \& {Gieles}}]{renaud17}
{Renaud}, F., {Agertz}, O., \& {Gieles}, M. 2017, \mnras, 465, 3622

\bibitem[{{Rom{\'a}n} \& {Trujillo}(2017)}]{roman17}
{Rom{\'a}n}, J. \& {Trujillo}, I. 2017, \mnras, 468, 703

\bibitem[{{Samurovi{\'c}}(2014)}]{samur14}
{Samurovi{\'c}}, S. 2014, \aap, 570, A132

\bibitem[{{S{\'a}nchez-Salcedo} {et~al.}(2006){S{\'a}nchez-Salcedo},
  {Reyes-Iturbide}, \& {Hernandez}}]{sanchezsalcedo06}
{S{\'a}nchez-Salcedo}, F.~J., {Reyes-Iturbide}, J., \& {Hernandez}, X. 2006,
  \mnras, 370, 1829

\bibitem[{{Sandage} \& {Binggeli}(1984)}]{sandage84}
{Sandage}, A. \& {Binggeli}, B. 1984, \aj, 89, 919

\bibitem[{{Skordis} \& {Z{\l}o{\'s}nik}(2019)}]{skordis19}
{Skordis}, C. \& {Z{\l}o{\'s}nik}, T. 2019, \prd, 100, 104013

\bibitem[{{Skordis} \& {Z{\l}o{\'s}nik}(2020)}]{skordis20}
{Skordis}, C. \& {Z{\l}o{\'s}nik}, T. 2020, arXiv e-prints, arXiv:2007.00082

\bibitem[{{Solanes} {et~al.}(2018){Solanes}, {Perea}, \&
  {Valent{\'\i}-Rojas}}]{solanes18}
{Solanes}, J.~M., {Perea}, J.~D., \& {Valent{\'\i}-Rojas}, G. 2018, \aap, 614,
  A66

\bibitem[{{Sparke} \& {Gallagher}(2000)}]{sparke00}
{Sparke}, L.~S. \& {Gallagher}, John~S., I. 2000, {Galaxies in the universe :
  an introduction}

\bibitem[{{Spitler} \& {Forbes}(2009)}]{spitler09}
{Spitler}, L.~R. \& {Forbes}, D.~A. 2009, \mnras, 392, L1

\bibitem[{{Spitzer}(1958)}]{spitzer58}
{Spitzer}, Lyman, J. 1958, \apj, 127, 17

\bibitem[{{Teyssier}(2002)}]{ramses}
{Teyssier}, R. 2002, \aap, 385, 337

\bibitem[{{Tiret} \& {Combes}(2007{\natexlab{a}})}]{tiret07}
{Tiret}, O. \& {Combes}, F. 2007{\natexlab{a}}, \aap, 464, 517

\bibitem[{{Tiret} \& {Combes}(2007{\natexlab{b}})}]{tiret07c}
{Tiret}, O. \& {Combes}, F. 2007{\natexlab{b}}, arXiv e-prints, arXiv:0712.1459

\bibitem[{{Toloba} {et~al.}(2018){Toloba}, {Lim}, {Peng}, {Sales},
  {Guhathakurta}, {Mihos}, {C{\^o}t{\'e}}, {Boselli}, {Cuillandre},
  {Ferrarese}, {Gwyn}, {Lan{\c{c}}on}, {Mu{\~n}oz}, \& {Puzia}}]{Toloba2018}
{Toloba}, E., {Lim}, S., {Peng}, E., {et~al.} 2018, \apjl, 856, L31

\bibitem[{{Tremaine} \& {Weinberg}(1984)}]{tremaine84}
{Tremaine}, S. \& {Weinberg}, M.~D. 1984, \mnras, 209, 729

\bibitem[{{Trujillo} {et~al.}(2019){Trujillo}, {Beasley}, {Borlaff},
  {Carrasco}, {Di Cintio}, {Filho}, {Monelli}, {Montes}, {Rom{\'a}n},
  {Ruiz-Lara}, {S{\'a}nchez Almeida}, {Valls-Gabaud}, \&
  {Vazdekis}}]{trujillo19}
{Trujillo}, I., {Beasley}, M.~A., {Borlaff}, A., {et~al.} 2019, \mnras, 486,
  1192

\bibitem[{{van der Burg} {et~al.}(2017){van der Burg}, {Hoekstra}, {Muzzin},
  {Sif{\'o}n}, {Viola}, {Bremer}, {Brough}, {Driver}, {Erben}, {Heymans},
  {Hildebrandt}, {Holwerda}, {Klaes}, {Kuijken}, {McGee}, {Nakajima},
  {Napolitano}, {Norberg}, {Taylor}, \& {Valentijn}}]{vanderburg17}
{van der Burg}, R. F.~J., {Hoekstra}, H., {Muzzin}, A., {et~al.} 2017, \aap,
  607, A79

\bibitem[{{van Dokkum} {et~al.}(2016){van Dokkum}, {Abraham}, {Brodie},
  {Conroy}, {Danieli}, {Merritt}, {Mowla}, {Romanowsky}, \&
  {Zhang}}]{vandokkum16}
{van Dokkum}, P., {Abraham}, R., {Brodie}, J., {et~al.} 2016, \apj, 828, L6

\bibitem[{{van Dokkum} {et~al.}(2018{\natexlab{a}}){van Dokkum}, {Cohen},
  {Danieli}, {Kruijssen}, {Romanowsky}, {Merritt}, {Abraham}, {Brodie},
  {Conroy}, {Lokhorst}, {Mowla}, {O'Sullivan}, \& {Zhang}}]{vandokkum18}
{van Dokkum}, P., {Cohen}, Y., {Danieli}, S., {et~al.} 2018{\natexlab{a}},
  \apjl, 856, L30

\bibitem[{{van Dokkum} {et~al.}(2018{\natexlab{b}}){van Dokkum}, {Danieli},
  {Cohen}, {Merritt}, {Romanowsky}, {Abraham}, {Brodie}, {Conroy}, {Lokhorst},
  {Mowla}, {O'Sullivan}, \& {Zhang}}]{2018Natur.555..629V}
{van Dokkum}, P., {Danieli}, S., {Cohen}, Y., {et~al.} 2018{\natexlab{b}},
  \nat, 555, 629

\bibitem[{{van Dokkum} {et~al.}(2015{\natexlab{a}}){van Dokkum}, {Abraham},
  {Merritt}, {Zhang}, {Geha}, \& {Conroy}}]{udgs}
{van Dokkum}, P.~G., {Abraham}, R., {Merritt}, A., {et~al.} 2015{\natexlab{a}},
  \apjl, 798, L45

\bibitem[{{van Dokkum} {et~al.}(2015{\natexlab{b}}){van Dokkum}, {Romanowsky},
  {Abraham}, {Brodie}, {Conroy}, {Geha}, {Merritt}, {Villaume}, \&
  {Zhang}}]{vandokkum15}
{van Dokkum}, P.~G., {Romanowsky}, A.~J., {Abraham}, R., {et~al.}
  2015{\natexlab{b}}, \apjl, 804, L26

\bibitem[{{Venhola} {et~al.}(2017){Venhola}, {Peletier}, {Laurikainen}, {Salo},
  {Lisker}, {Iodice}, {Capaccioli}, {Verdois Kleijn}, {Valentijn}, {Mieske},
  {Hilker}, {Wittmann}, {van de Ven}, {Grado}, {Spavone}, {Cantiello},
  {Napolitano}, {Paolillo}, \& {Falc{\'o}n-Barroso}}]{venhola17}
{Venhola}, A., {Peletier}, R., {Laurikainen}, E., {et~al.} 2017, \aap, 608,
  A142

\bibitem[{{Villegas} {et~al.}(2010){Villegas}, {Jord{\'a}n}, {Peng},
  {Blakeslee}, {C{\^o}t{\'e}}, {Ferrarese}, {Kissler-Patig}, {Mei}, {Infante},
  {Tonry}, \& {West}}]{villegas10}
{Villegas}, D., {Jord{\'a}n}, A., {Peng}, E.~W., {et~al.} 2010, \apj, 717, 603

\bibitem[{{Wittenburg} {et~al.}(2020){Wittenburg}, {Kroupa}, \&
  {Famaey}}]{wittenburg20}
{Wittenburg}, N., {Kroupa}, P., \& {Famaey}, B. 2020, \apj, 890, 173

\bibitem[{{Yagi} {et~al.}(2016){Yagi}, {Koda}, {Komiyama}, \&
  {Yamanoi}}]{yagi16}
{Yagi}, M., {Koda}, J., {Komiyama}, Y., \& {Yamanoi}, H. 2016, \apjs, 225, 11

\bibitem[{{Zaritsky} {et~al.}(2019){Zaritsky}, {Donnerstein}, {Dey},
  {Kadowaki}, {Zhang}, {Karunakaran}, {Mart{\'\i}nez-Delgado}, {Rahman}, \&
  {Spekkens}}]{zaritsky19}
{Zaritsky}, D., {Donnerstein}, R., {Dey}, A., {et~al.} 2019, \apjs, 240, 1

\bibitem[{{Zhao} {et~al.}(2013){Zhao}, {Famaey}, {L{\"u}ghausen}, \&
  {Kroupa}}]{zhao13}
{Zhao}, H., {Famaey}, B., {L{\"u}ghausen}, F., \& {Kroupa}, P. 2013, \aap, 557,
  L3

\end{thebibliography}

\end{document}